\documentclass[aps,prl,twocolumn,amsmath,amssymb,superscriptaddress]{revtex4-2}

\usepackage[utf8]{inputenc}
\usepackage[french]{babel}

\usepackage{amsmath}
\usepackage{amssymb}
\usepackage{braket}
\usepackage{graphicx}
\usepackage{color}
\usepackage{setspace}
\usepackage[dvipsnames]{xcolor}
\usepackage{ulem} 

\newcommand{\LL}[1]{\textcolor{black}{#1}}

\begin{document}

	
	
	
	
	\title{Controlling photon polarisation with a single quantum dot spin}
	
	
	

	\author{E. Mehdi}
	\affiliation{Centre for Nanosciences and Nanotechnology, CNRS, Universit\'e Paris-Saclay, UMR 9001,
		10 Boulevard Thomas Gobert, 91120, Palaiseau, France}
	\affiliation{Universit\'e Paris Cit\'e, Centre for Nanoscience and Nanotechnology, F-91120 Palaiseau, France}
	\author{M. Gundin-Martinez}
	\affiliation{Centre for Nanosciences and Nanotechnology, CNRS, Universit\'e Paris-Saclay, UMR 9001,
		10 Boulevard Thomas Gobert, 91120, Palaiseau, France}
	\author{C. Millet}
	\affiliation{Centre for Nanosciences and Nanotechnology, CNRS, Universit\'e Paris-Saclay, UMR 9001,
		10 Boulevard Thomas Gobert, 91120, Palaiseau, France}
	\author{N. Somaschi}
	\affiliation{Centre for Nanosciences and Nanotechnology, CNRS, Universit\'e Paris-Saclay, UMR 9001,
		10 Boulevard Thomas Gobert, 91120, Palaiseau, France}
	\affiliation{Quandela SAS, 10 Boulevard Thomas Gobert, 91120, Palaiseau, France}
	\author{A. Lemaître}
	\affiliation{Centre for Nanosciences and Nanotechnology, CNRS, Universit\'e Paris-Saclay, UMR 9001,
		10 Boulevard Thomas Gobert, 91120, Palaiseau, France}
	\author{I. Sagnes}
	\affiliation{Centre for Nanosciences and Nanotechnology, CNRS, Universit\'e Paris-Saclay, UMR 9001,
		10 Boulevard Thomas Gobert, 91120, Palaiseau, France}
	\author{L. Le Gratiet}
	\affiliation{Centre for Nanosciences and Nanotechnology, CNRS, Universit\'e Paris-Saclay, UMR 9001,
		10 Boulevard Thomas Gobert, 91120, Palaiseau, France}
	\author{D. Fioretto}
	\affiliation{Centre for Nanosciences and Nanotechnology, CNRS, Universit\'e Paris-Saclay, UMR 9001,
		10 Boulevard Thomas Gobert, 91120, Palaiseau, France}
	\author{N. Belabas}
	\affiliation{Centre for Nanosciences and Nanotechnology, CNRS, Universit\'e Paris-Saclay, UMR 9001,
		10 Boulevard Thomas Gobert, 91120, Palaiseau, France}
	\author{O. Krebs}
	\affiliation{Centre for Nanosciences and Nanotechnology, CNRS, Universit\'e Paris-Saclay, UMR 9001,
		10 Boulevard Thomas Gobert, 91120, Palaiseau, France}
	\author{P. Senellart}
	\affiliation{Centre for Nanosciences and Nanotechnology, CNRS, Universit\'e Paris-Saclay, UMR 9001,
		10 Boulevard Thomas Gobert, 91120, Palaiseau, France}
	\author{L. Lanco}
	\affiliation{Centre for Nanosciences and Nanotechnology, CNRS, Universit\'e Paris-Saclay, UMR 9001,
		10 Boulevard Thomas Gobert, 91120, Palaiseau, France}
	\affiliation{Universit\'e Paris Cit\'e, Centre for Nanoscience and Nanotechnology, F-91120 Palaiseau, France} 
	\affiliation{Institut Universitaire de France (IUF)}
	
	
	
	\begin{abstract}
		In the framework of optical quantum computing and communications, a major objective consists in building receiving nodes that implement conditional operations on incoming photons, using the interaction with a single stationary qubit. In particular, the quest for scalable nodes motivated the development of  cavity-enhanced spin-photon interfaces with solid-state emitters. An important challenge remains, however, to produce a stable, controllable, spin-dependant photon state, in a deterministic way.
		Here we use a pillar-based high-Q cavity, embedding a singly-charged semiconductor quantum dot, to demonstrate the control of giant polarisation rotations induced by a single electron spin. A complete tomography approach is used to deduce the output polarisation Stokes vector, conditioned by a single spin state.  
		We experimentally demonstrate rotation amplitudes such as $\pm \frac{\pi}{2}$ and $\pi$ in the Poincaré sphere, as required for applications based on spin-polarisation mapping and spin-mediated photon-photon gates. In agreement with our modeling, we observe that the environmental noise does not limit the amplitude of the spin-induced rotation, yet slightly degrades the polarisation purity of the output states. We find that the polarisation state of the reflected photons can be manipulated in most of the Poincaré sphere, through controlled spin-induced rotations, thanks to  moderate cavity birefringence and limited noise. This control allows the operation of spin-photon interfaces in various configurations, including at zero or low magnetic fields, which ensures compatibility with key protocols for photonic cluster state generation.
	\end{abstract}

	\maketitle
	
	\textbf{Introduction.}
	A major challenge for optical quantum information
	is the development of deterministic light-matter interfaces, used as stationary nodes communicating through photons \cite{Cirac1997}
	\LL{Potentially, efficient quantum communication and quantum computing could be performed in a loss-resistant way, with only a few nodes \cite{Borregaard2020}, or even a single one \cite{Pichler2017,Sun2020,Shi2021}, used to emit and receive photons.}
	
	\LL{In the last decade, important efforts have thus been devoted to build efficient receiving nodes, performing conditional operations on incoming photons, using cavity-QED devices.
		This led, for instance, to the demonstration of various quantum gates between incoming photons and stationary qubits, using e.g. atoms \cite{Reiserer2014,Tiecke2014,Shomroni2014,Bechler2018}, solid-state spins \cite{Kim2013,Sun2016,Nguyen2019,Chan2022,Antoniadis2022}, and superconducting qubits \cite{Reuer2022}. The developed interfaces could subsequently be used in various demonstrations of photon-photon gates \cite{Hacker2016,Stolz2022}, single-photon transistors \cite{Shomroni2014, Sun2018, Wang2022}, quantum memories \cite{Bhaskar2020}, and quantum non-demolition detectors \cite{Kono2018}.}
	
	\LL{Within this quest, interfacing solid-state spins with optical photons holds important promises in terms of scalability and long-distance communications. This includes a number of possible emitters and cavity structures  \cite{Atature2018,Awschalom2018}, as well as various combinations of photonic encodings, including polarisation,  path, and time-bin \cite{Kim2013,Sun2016,Nguyen2019,Chan2022,Antoniadis2022}.}
	
	\LL{Polarisation encoding offers, in particular, the most direct approach to realize deterministic spin-photon and multi-photon gates  \cite{Leuenberger2006, Hu2008,Lindner2009,Chen2021}, together with straightforward 1-qubit gates and measurements. A key objective is to produce a perfect spin-polarisation mapping, where opposite spin states $\ket{\uparrow}$ and $\ket{\downarrow}$ map to opposite polarisation states for the interfaced photon, $\ket{\Psi_{\uparrow}}$ and $\ket{\Psi_{\downarrow}}$, verifying $\langle \Psi_{\uparrow} | \Psi_{\downarrow} \rangle=0$. In this respect, most realizations have been pioneered with highly-birefringent structures \cite{Kim2013,Tiecke2014,Sun2016}, and at high magnetic fields. In such a case, the protocols aim at exploiting the  spin-dependent $\pi$-phase shift that, in the limit of an ideal device \cite{Duan2004}, would be induced on the reflected or transmitted photons.}

	
	\LL{Alternatively, a promising strategy is to use cavity-QED devices with low birefringence, as already implemented using semiconductor quantum dots (QDs) in pillar-based structures \cite{Rakher2009,Arnold2015, Androvi2019,Wells2019}. 
		Pillar cavities indeed allow a robust and deterministic light-matter coupling compatible with electrical control \cite{Nowak2014}. 
		They enable the efficient injection \cite{Hilaire2018} and extraction  \cite{Somaschi2016,Ding2016} of photons, their deterministic interaction with the embedded qubit \cite{DeSantis2017}, as well as a moderate birefringence, i.e. a spectral overlap between orthogonally-polarised cavity modes \cite{Rakher2009,Hilaire2018,Ding2016}. 
		This could lead to perfect spin-polarisation mapping through $\pm \frac{\pi}{2}$ spin-dependant rotations, including at zero or low magnetic field, to ensure compatibility with key protocols for the generation of photonic cluster states \cite{Lindner2009}. Yet, until now, the spin-induced polarisation rotations have remained limited by optical losses \cite{Arnold2015} and/or by detrimental noise \cite{Arnold2015, Wells2019,Androvi2019}.}
	
	\LL{Interestingly, in the variety of polarisation-based experiments  \cite{Tiecke2014,Kim2013,Sun2016,Rakher2009,Arnold2015, Androvi2019,Wells2019} spin-induced polarisation rotations have mostly been translated into measured intensity contrasts for a given basis. 
		This is sufficient to characterize the fidelity of a polarisation state with respect to a specific ideal target, yet does not give access to the direction and amplitude of the polarisation rotation in the Poincaré sphere, nor to the purity of the produced state. 
		This lack of knowledge, regarding the polarisation states experimentally produced, makes it difficult to differentiate between  technological and experimental limitations, and mitigate them with adapted protocols. In addition, an open question remained regarding the range of achievable polarisation states that can be produced with a given device. }

	Here, we report on the observation and control of giant polarisation rotations induced by a single QD-embedded electron spin, deterministically coupled to an electrically-contacted pillar cavity device \LL{(see Methods).
		The Purcell enhancement provided by the high-Q cavity, together with the \LL{low level} of charge noise
		, allow reaching 
		giant polarisation rotations\LL{, including the highly-desired configurations of $\pm \frac{\pi}{2}$ and full $\pi$ rotations}.
		We use polarisation tomography} \cite{Anton2017} to fully characterise the state of the reflected photons in the Poincaré sphere. The possibility to add or remove the \textcolor{black}{electron} from the quantum dot, through the applied bias, then allows deducing the conditional Stokes vector $\vec{S}_{\uparrow}$, conditioned to a charged QD in the spin state $\uparrow$. The experimental data are in good agreement with theoretical simulations:  
	we find, in particular, that the residual noise does not limit the amplitude of the polarisation rotation, yet limits the polarisation purity of the output state. 
	We finally demonstrate that, by a proper set of detunings, any orientation of the output Stokes vector  $\vec{S}_{\uparrow}$ can be reached. We find this possible, in particular, thanks to the moderate cavity birefringence, which allows Purcell-enhancing the light-matter interaction in all polarisations. Accessing arbitrary latitudes and longitudes in the Poincaré sphere provides essential degrees of freedom for the perfect mapping of spin states into opposite polarisation states, to ensure the adaptability of devices to a wide range of protocols and experimental conditions.\\
	

	\begin{figure*}
		\includegraphics[scale=0.22]{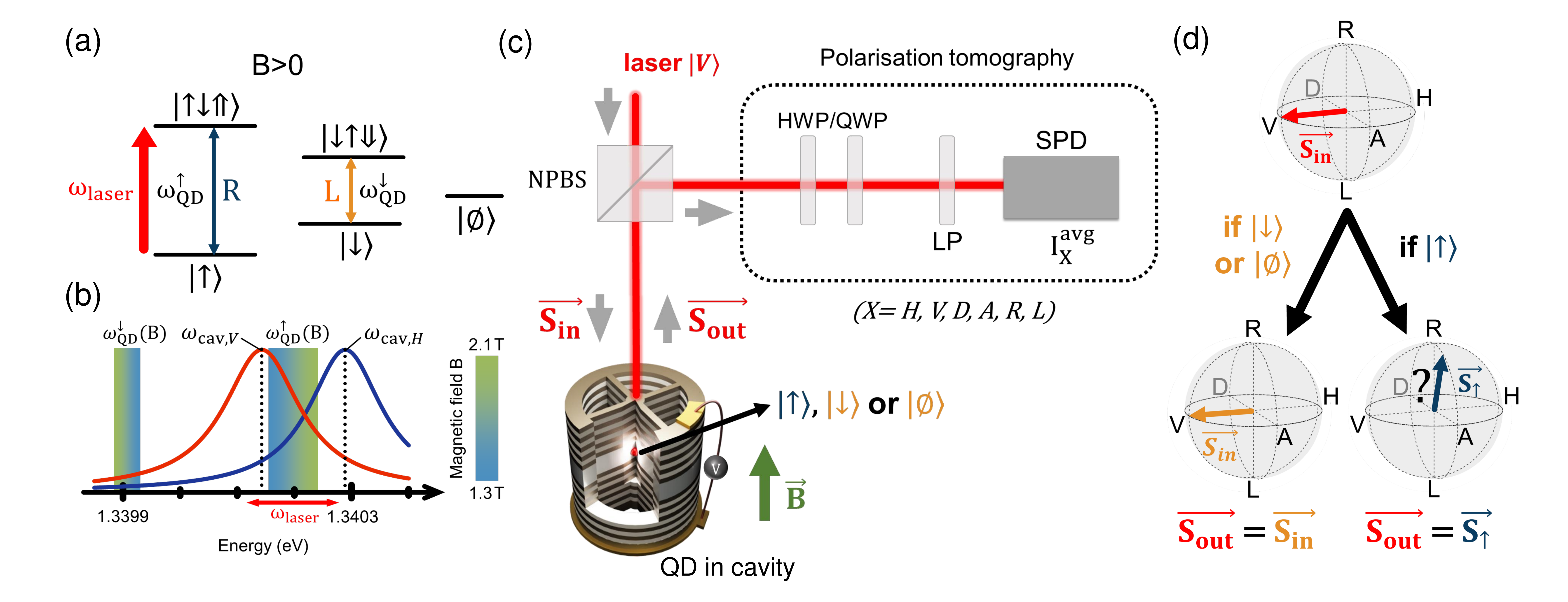}
		\caption{\textbf{Spin-selective polarisation rotation} - 
			\textbf{(a)}
			Energy levels of the negatively charged quantum dot, with an applied magnetic field B (Faraday configuration).
			The system is probed with a laser of energy $\omega_{\mathrm{laser}}$ close to $\omega_{\mathrm{QD}}^{\uparrow}$. 
			\textbf{(b)}
			Scale comparing the two transitions energies, $\omega_{\mathrm{QD}}^{\uparrow}$ and $\omega_{\mathrm{QD}}^{\downarrow}$, to the $V$ and $H$ cavity modes energies, $\omega_{\mathrm{cav},V} = 1.340143\,eV$ and $\omega_{\mathrm{cav},H} = 1.340289\,eV$.
			The energies $\omega_{\mathrm{QD}}^{\uparrow}$ and $\omega_{\mathrm{QD}}^{\downarrow}$ depend on the value of B, as shown by the colorscale,
			\textcolor{black}{ and the two cavity modes spectrally overlap.}
			\textbf{(c)} 
			Experimental scheme. 
			The incoming laser, of Stokes vector $\vec{S}_{\mathrm{in}}  \equiv \ket{V}$, is sent on an electrically contacted quantum dot – micropillar device. 
			The reflected polarisation state $\vec{S}_{\mathrm{out}} $ is analysed through a polarisation tomography setup (dashed box) measuring the average reflected intensity $I_X^{\mathrm{avg} }$ along the six polarisations $X =$ $H$, $V$, $D$, $A$, $R$, $L$. 
			NPBS: Non-Polarising Beamsplitter – H(Q)WP: Half (Quarter) Wave Plate – LP: Linear Polariser – SPD: Single Photon Detector. 
			\textbf{(d)} 
			The possible output polarisations of the reflected light in the Poincaré sphere. 
		}
		\centering
	\end{figure*}

	\textbf{Principle of the experiments.}
	We sketch in Fig.  1a the charged QD energy levels, with two ground states with opposite electron spin  $\ket{\uparrow}$ and $\ket{\downarrow}$, and their corresponding trion states $\ket{\uparrow \downarrow \Uparrow }$ and $\ket{\downarrow \uparrow \Downarrow}$, consisting in a pair of electrons and a single hole \cite{Hilaire2020, Warburton2013}. 
	The electron might escape the QD, as described by the additional empty state denoted $\ket{\varnothing}$ \cite{Mannel2021}.  
	Here, we work in the particular configuration, 
	where an external longitudinal magnetic field  B (Faraday configuration) is applied to the QD, lifting the energy degeneracy between the two transitions through the Zeeman effect \cite{Urbaszek2013}.
	For a magnetic field around $2\,T$, the two transitions have no energy overlap: we respectively label $\omega_{\mathrm{QD}}^{\uparrow}$ and $\omega_{\mathrm{QD}}^{\downarrow}$ the  $\ket{\uparrow} - \ket{\uparrow\downarrow \Uparrow}$ and $\ket{\downarrow} - \ket{\downarrow \uparrow \Downarrow}$ transition energies (in $\hbar=1$ units), with $\omega_{\mathrm{QD}}^{\uparrow} > \omega_{\mathrm{QD}}^{\downarrow}$. 
	A tunable narrow-band laser of energy $\omega_{\mathrm{laser}}$ close to $\omega_{\mathrm{{QD}}}^{\uparrow}$ selectively probes the $\ket{\uparrow} - \ket{\uparrow\downarrow \Uparrow}$ transition while the detuned $\ket{\downarrow} - \ket{\downarrow \uparrow \Downarrow}$ transition remains negligibly excited (see Methods).
	
	To provide an efficient spin-photon interface, the QD is deterministically coupled to a pillar-based, electrically contacted microcavity \cite{Nowak2014}. Furthermore, the transition energy $\omega_{\mathrm{{QD}}}^{\uparrow}$ is varied, with the applied magnetic field, in the vicinity of the two cavity mode resonances $\omega_{\mathrm{cav},H}$ and $\omega_{\mathrm{cav},V}$, as displayed in Fig. 1b. 
	This ensures that the $\ket{\uparrow} - \ket{\uparrow\downarrow \Uparrow}$ transition, at $\omega_{\mathrm{{QD}}}^{\uparrow}(B)$, benefits from an efficient Purcell enhancement.
	
	The experimental scheme is sketched in Fig. 1c. 
	The incoming polarisation state, whose Stokes vector is denoted $\vec{S}_{\mathrm{in}} $, is adjusted to match one of the two cavity polarisation eigenaxes, the direction of which defines the vertical polarisation $V$. 
	By doing so, we ensure that there is no cavity-induced polarisation rotation \cite{Hilaire2018}.
	A non-polarising beamsplitter directs the reflected light, whose polarisation Stokes vector is denoted $\vec{S}_{\mathrm{out}} $, to the polarisation tomography setup \cite{Hilaire2018, Anton2017} that measures the reflected intensities 
	$I_X^{\mathrm{avg}}$, in various polarisation states $X=H,V,D,A,R,L$ (respectively corresponding to linear horizontal, vertical, diagonal and antidiagonal polarisations, and circular right- and left-handed polarisations).

	The notation $I_X^{\mathrm{avg}}$ denotes the average reflected intensity in polarisation $X$. 
	$I_X^{\mathrm{avg}}$ is measured with a single photon detector, by integrating counts for $0.1\,s$,
	a timescale multiple orders of magnitude higher than the characteristic times of electron spin-flip and escape/capture processes. 
	During this integration time, as the spin is not initialised, we thus measure the average optical response over the possible ground states $\ket{\uparrow}$, $\ket{\downarrow}$ and $\ket{\varnothing}$.

	If the QD is in state  $\ket{\downarrow}$ or $\ket{\varnothing}$, as illustrated in Fig. 1d, the reflected polarisation state is unchanged ($\vec{S}_{\mathrm{out}}  = \vec{S}_{\mathrm{in}}$), due to the absence of QD-induced rotation. 
	When the ground state is $\ket{\uparrow}$, on the contrary, the spin-photon interface can induce large polarisation rotations (Kerr rotation), leading to a different output Stokes vector $\vec{S}_{\mathrm{out}}=\vec{S}_{\uparrow}$. 
	In the absence of environmental noise, and at low excitation power, $\vec{S}_{\uparrow}$ corresponds to a pure polarisation state, $|\vec{S}_{\uparrow}| =1$, at the surface of the sphere. 
	The input polarisation $\vec{S}_{\mathrm{in}}  \equiv \ket{V}  \propto \ket{R} - \ket{L} $ is indeed converted to $\vec{S}_{\uparrow} \equiv \ket{\psi_{\uparrow}}	\propto r_R^{\uparrow}\ket{R} - r_L^{\uparrow}\ket{L}$, where $r_R^{\uparrow}$
	and $r_L^{\uparrow}$ are the complex reflection coefficients for $R$-polarised and $L$-polarised light, when the ground state is $\ket{\uparrow}$. 
	In the presence of environmental noise, a similar rotation of polarisation is expected, yet with a potential degradation of the polarisation purity \cite{Anton2017}.  \\

	\begin{figure*}
		\centering
		\includegraphics[scale=0.20]{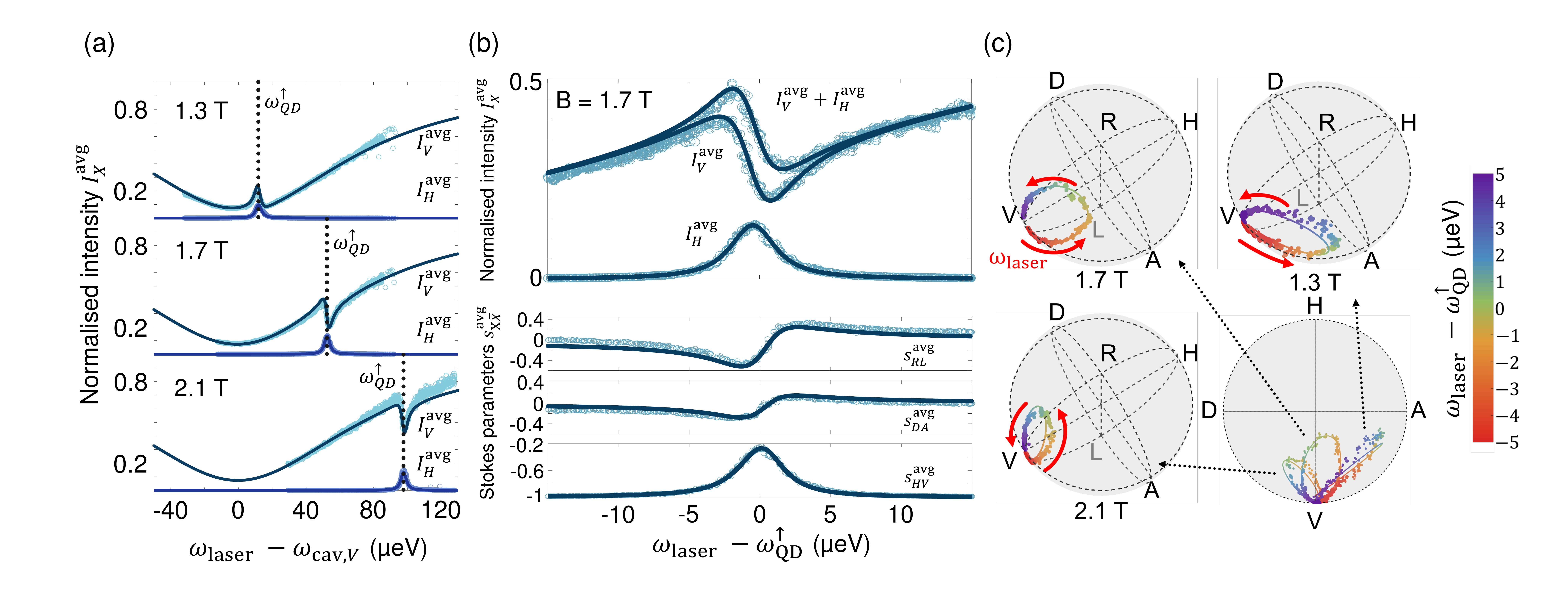}
		\caption{ \textbf{Averaged intensities and polarisations for a non initialised} QD state - (circles: experimental data; lines: numerical simulations)
			\textbf{(a)}
			Normalised average intensity $I_X^{\mathrm{avg}}$ (with X=H, V) as a function of the energy detuning between the laser and the V cavity mode for three different magnetic fields $1.3\, \mathrm{T}$, $1.7\, \mathrm{T}$ and $2.1\, \mathrm{T}$. 
			The measured signal results from the averaging over QD states ($\ket{\uparrow}$, $\ket{\downarrow}$ and $\ket{\varnothing}$), inherent to the measurement timescale. 
			The energy $\omega_{\mathrm{QD}}^{\uparrow}$ of the $\ket{\uparrow}$-$\ket{\uparrow \downarrow \Uparrow }$ transition is shifted by the applied magnetic field (see Fig. 1b). 
			\textbf{(b) }
			Normalised average intensity $I_X^{\mathrm{avg}}$ (with X=H, V) and Stokes parameters $s_{X\bar{X}}^{\mathrm{avg}}$ (with $X\bar{X}$=HV, DA, RL) as a function of the energy detuning between the laser and the $\ket{\uparrow} - \ket{\uparrow \downarrow \Uparrow }$ transition at $1.7 \,$T. 
			\textbf{(c) }
			Rotation of the average output polarisation in the Poincaré sphere as a function of the detuning (see colorscale) between the laser and the $\ket{\uparrow} - \ket{\uparrow \downarrow \Uparrow }$ transition, for the three different magnetic fields. 
			The bottom-right panel displays the output polarisation in a top view of the Poincaré sphere for the three magnetic fields. 
			The red arrows describe increasing laser-QD detunings.
		}
	\end{figure*} 
	
	\textbf{Results.}
	We start by measuring the reflected light, averaged over the non-initialised QD state, in the six polarisation bases. 
	In particular, Fig. 2a displays the reflected light intensities $I_{H/V}^{\mathrm{avg}}$, normalised by the input laser intensity, and plotted as a function of the energy detuning between the laser and the cavity mode $V$ for three different magnetic fields. 
	The applied magnetic field controls the splitting between the two transitions of energy $\omega_{\mathrm{QD}}^{\uparrow}$ and $\omega_{\mathrm{QD}}^{\downarrow}$, bringing $\omega_{\mathrm{QD}}^{\uparrow}$ in and out of resonance with the cavity mode $V$ of energy $\omega_{\mathrm{cav}, V}$ (see Fig. 1b). 
	The transition at energy $\omega_{\mathrm{QD}}^{\downarrow}$ remains detuned from both cavity modes, and outside the spectral range probed in Fig. 2a.

	The incoming intensity being V-polarised, $I_H^{\mathrm{avg}}$ corresponds only to the QD resonance fluorescence (RF) emission, cross-polarised to the incoming laser polarisation.
	By contrast the reflected intensity $I_V^{\mathrm{avg} }$ results from the interference between the empty cavity reflectivity, contributing to the reflectivity dip centered at $\omega_{\mathrm{cav}, V}$, and the co-polarised part of the QD RF emission, contributing to the optical response in the vicinity of $\omega_{\mathrm{QD}}^{\uparrow}$.
	
	Focusing on a smaller energy range around $\omega_{\mathrm{QD}}^{\uparrow}$, at $1.7\, \mathrm{T}$, the top panel of Fig. 2b shows the dependence of the reflected intensities, and their sum $I_{V}^{\mathrm{avg} } + I_H^{\mathrm{avg}}$, on the energy detuning between the laser and $\omega^{\uparrow}_{\mathrm{QD}}$. 
	In the lower panels, the Stokes parameters $s_{X\bar{X}}^{\mathrm{avg}}$ of the reflected light, retrieved via full tomography (see Methods), are displayed. 
	Far from the resonance with the $\omega^{\uparrow}_{\mathrm{QD}}$ transition, the reflected polarisation is identical to the input one: $s_{HV}^{\mathrm{avg} }$ = -1, $s_{DA}^{\mathrm{avg}}$ = $s_{RL}^{\mathrm{avg}}$  = 0, $\vec{S}_{\mathrm{out}} $ = $\vec{S}_{\mathrm{in}} \equiv \ket{V}$.  
	Conversely, the laser in resonance with $\omega^{\uparrow}_{\mathrm{QD}}$ results in a maximum for $s_{HV}^{\mathrm{avg}}$, consistent with the observed peak in the cross-polarised resonance fluorescence $I_H^{\mathrm{avg}}$. 
	

	The experimental data for both the intensities and Stokes vectors are in good agreement with the numerical simulations. 
	Simulations are carried out with a custom-made cavity quantum electrodynamics (QED) platform \cite{Arnold2015}, built on the Quantum Optics Toolbox \cite{Tan1999}, to solve the Master equation describing the coupled QD-cavity system. 
	We use the input-output formalism as detailed in the Supplemental section 3. 
	The simulated behaviour of the cavity is described by its two modes $H$ and $V$ with their energies $\omega_{\mathrm{cav}, V}$ and $\omega_{\mathrm{cav}, H}$, and their damping rates $\kappa_V = (162 \pm 6)\, \mu eV$ and $\kappa_H = (155 \pm 6)\, \mu eV$ that govern the rate at which H-polarised and V-polarised photons escape the cavity. 
	$\omega_{V/H}$ and $\kappa_{V/H}$ are deduced from the empty cavity reflectivity spectrum (See Supplemental section 4).  
	The splitting between the two cavity modes is given by $\Delta = \omega_{\mathrm{cav}, H} -\omega_{\mathrm{cav}, V}  = (146 \pm 1)\, \mu eV$, of the order of $\kappa_H$ and $\kappa_V$. 
	This characterizes a partial overlap between the modes, as schematized in Fig. 1b. 
	The probability that an intracavity photon escapes from the top mirror of the micropillar and couples out to the experimental setup is given by the effective top mirror output coupling $\eta_{\mathrm{top, eff}} = (0.635 \pm 0.005)$ (see Supplemental section 4).
	The coupling between an intracavity photon and the QD is described by the light-matter coupling constant of $g =(15.0 \pm 0.5) \, \mu eV $. 
	We also consider the spontaneous decay rate $\gamma_{sp} = (0.35 \pm 0.05)\, \mu eV$ to take into account the emission in all spatial modes other than the two fundamental cavity modes $H$ and $V$.
	$g$ and $\gamma_{sp}$ govern both the spectral width and the amplitude of the cross-polarised QD RF emission, $I_H^{\mathrm{avg} }$.
	These parameters translate into a good cooperativity $C=\frac{2g^2}{\kappa \gamma_{sp}}=(8 \pm 2)$, in a model where pure dephasing is neglected.
	This value is comparable with reported values in electrically contacted QD-cavity devices embedding neutral QDs \cite{Somaschi2016}. 
	However, charge tunnelling and environmental noise tend to slightly degrade the device performance, as discussed in the following.
	
	The contributions of the states $\ket{\uparrow}$, $\ket{\downarrow}$ or $\ket{\varnothing}$ to the reflected light are analysed by introducing the associated occupation probabilities (resp. $P_{\uparrow}$, $P_{\downarrow}$ and $P_{\varnothing}$). 
	The charge occupation $P_c$, i.e. the probability for the quantum dot to be charged with a single electron, is $P_c = P_{\uparrow} + P_{\downarrow}$; consequently the empty state has an occupation $P_{\emptyset}$ of $1-P_c$. 
	In our electrically-contacted devices, $P_c$ is determined by co-tunnelling processes: when an electron tunnels out of the QD, a different electron can tunnel in from the electron reservoir, erasing all information about the spin state. 
	We can therefore consider that $P_{\uparrow} = P_{\downarrow}= \frac{P_c}{2}$. Under this assumption, we obtain a good fit of the measurements with $P_c = (94 \pm 6) \%$. 
	Finally, we also take into account the fluctuations of $\omega_{\mathrm{QD}}^{\uparrow}$ that are expected, in particular, due to the hyperfine interaction between the electron spin and the approximately $10^5$ QD nuclei. 
	In our model, those fluctuations are described by a Gaussian distribution of $\omega_{\mathrm{QD}}^{\uparrow}$, with a standard deviation $\sigma = (0.5 \pm 0.3 ) \mu eV$ in agreement with the hyperfine-induced noise observed with similar QDs \cite{Zhukov2018}. 
	We emphasize that the value of $\sigma$ can also include some contribution of charge noise \cite{Kuhlmann2013}, which would similarly lead to fluctuations of $\omega_{\mathrm{QD}}^{\uparrow}$ during the acquisition time (see Supplemental section 6).

	The polarisation rotation induced by the device is shown in Fig. 2c, in which the average Stokes vectors $\vec{S}_{\mathrm{out}} $ are plotted as a function of the detuning between the laser and $\omega_{\mathrm{QD}}^{\uparrow}$, for the different magnetic fields.
	The output polarisation $\vec{S}_{\mathrm{out}} $ is deduced from the six average reflected intensities $I_{X}^{\mathrm{avg}}$
	. Each point is color-coded to a specific detuning $\omega_{\mathrm{laser}} - \omega_{\mathrm{QD}}^{\uparrow}$.
	The top and bottom-left panels shows the polarisation rotation for the three magnetic fields $1.3\, \mathrm{T}$, $1.7\, \mathrm{T}$ and $2.1\, \mathrm{T}$, while the bottom right panel aggregates the three cases as viewed from the top of the Poincaré sphere (R facing up). 
	On resonance with the $\ket{\uparrow} - \ket{\uparrow \downarrow \Uparrow}$ transition (green data points), the output polarisation  is the farthest from the $\ket{V}$ input polarisation. Far from resonance, $\vec{S}_{\mathrm{out}}$ remains $\ket{V}$.
	The trajectory of $\vec{S}_{\mathrm{out}} $, ie. the ensemble of points in the Poincaré sphere as the laser wavelength is scanned, depends directly on the detuning between the $\ket{\uparrow} - \ket{\uparrow \downarrow \Uparrow}$ transition and the cavity modes H and V.
	Such a detuning is controlled by the applied magnetic field as shown in Fig. 2a. 
	The numerical simulations in solid lines matches the measured trajectory of the average output polarisation $\vec{S}_{\mathrm{out}} $, for all three magnetic fields, with the above-described set of parameters. 
	In particular, the limited noise at $\sigma = (0.5 \pm 0.3 ) \, \mu eV$, compared to the homogeneous, Purcell-accelerated linewidth of the QD transition ($\approx 3.5 \, \mu eV$, see Supplemental section 1) allows preserving large polarisation rotations in the Poincaré sphere. 

	\begin{figure}[h!]
		\centering
		\includegraphics[scale=0.22]{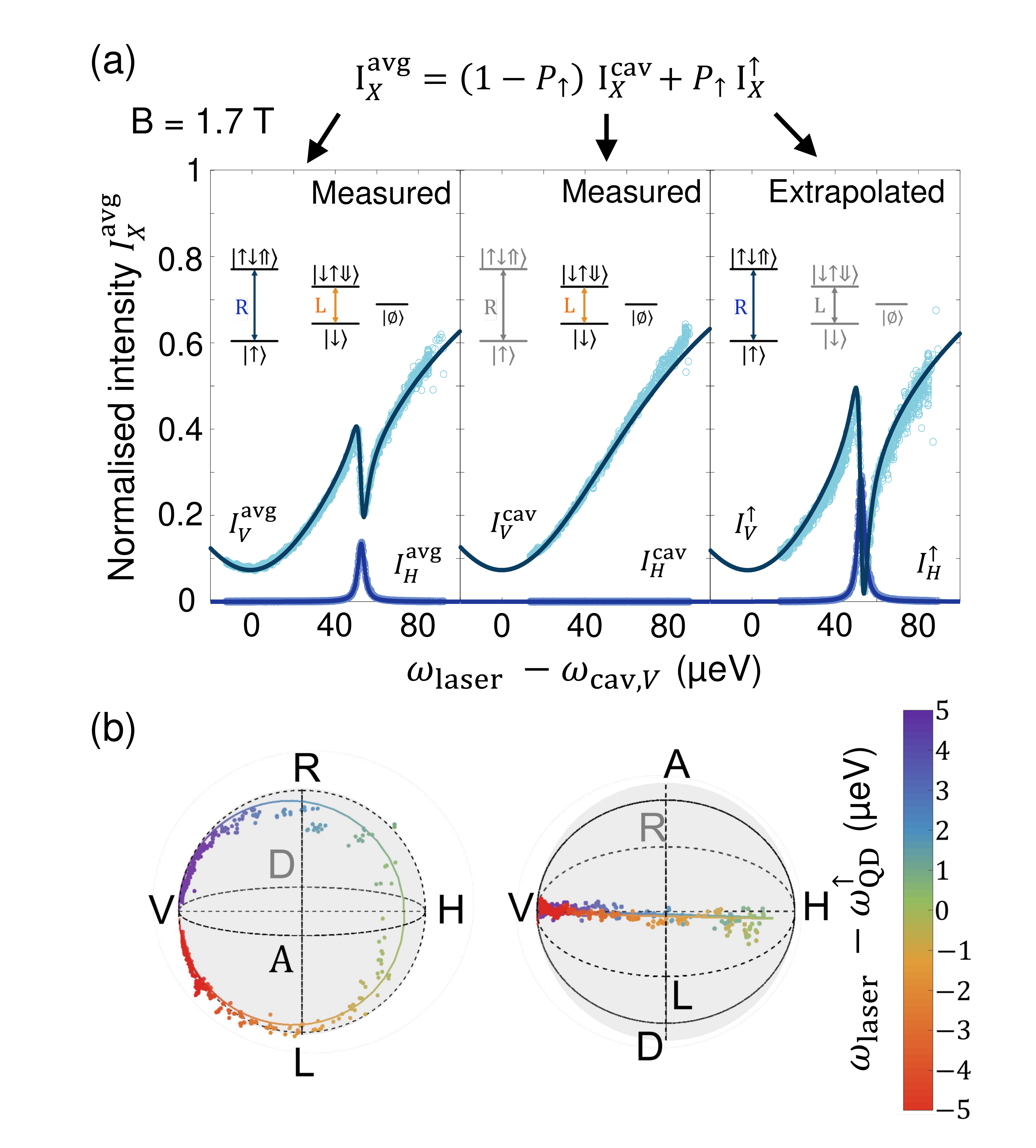}
		\caption{
			\textbf{Extrapolating the Stokes vector for a pure QD spin state } - 
			\textbf{(a)} 
			Normalised intensity $I_X^{\mathrm{avg} }$ (with X=H, V) as a function of the detuning between the laser and the cavity V-mode at $1.7~$T (circles: measured or extrapolated data; lines: numerical simulations). 
			First panel: experimentally measured intensity $I_X^{\mathrm{avg} }$ averaged over the non-initialised QD state ($\ket{\uparrow}$, $\ket{\downarrow}$ and $\ket{\varnothing}$), as in Fig. 2a middle panel.
			Second panel: experimentally measured intensity $I_X^{\mathrm{cav}}$ for the empty cavity, providing information on the system response when it is not in state $\ket{\uparrow}$. 
			Third panel: extrapolated intensity $I_X^{\uparrow}$ describing the system response when the QD is in state $\ket{\uparrow}$.
			\textbf{(b)} 
			Extrapolated Stokes vector $\vec{S_{\uparrow}}$ as a function of $\omega_{\mathrm{laser}} - \omega_{\mathrm{QD}}^{\uparrow}$ (see colorscale), viewed from two different angles in the Poincaré sphere (circles: extrapolated from experimental data; lines: numerical simulations). 
		}
	\end{figure}
	
	We now deduce the behaviour of the output polarisation $\vec{S}_{\uparrow}$, conditioned to a spin being in state $\ket{\uparrow}$, even though the electron spin is not experimentally initialised. 
	It can indeed be extrapolated from the measured average intensities and  complementary measurements with the empty cavity.
	When the QD is in the state $\ket{\uparrow}$ with a probability $P_{\uparrow}$
	, the reflected light is described by the set of intensities $I_X^{\uparrow}$, corresponding to the desired Stokes vector $\vec{S_{\uparrow}}$.
	Whereas, when the QD is in the state $\ket{\downarrow}$ or $\ket{\varnothing}$, with a probability $(1-P_{\uparrow})$, it is transparent for the laser at the studied energy range (see Fig. 1b and 1d). 
	The corresponding reflected light polarisation is then described by the empty cavity intensities $I_X^{\mathrm{cav}}$, that can be experimentally measured by forcing the absence of an electron in the QD (applied voltage of $0\, V$). 
	For each polarisation $X$
	, $I_X^{\mathrm{avg} }$ is thus the weighted sum of the conditional intensities in the two previous cases:
	\begin{equation}
		I_X^{\mathrm{avg} } = (1 - P_{\uparrow})I_X^{\mathrm{cav}} + P_{\uparrow} I_X^{\uparrow}.
	\end{equation}
	
	Fig. 3a illustrates the extrapolation process at $1.7\, \mathrm{T}$. 
	The measured normalised intensities $I_X^{\mathrm{avg} }$ and $I_X^{\mathrm{cav}}$ (for $X$ = $H$ and $X$ = $V$), are plotted as a function of the energy detuning between the laser and the cavity mode $V$ in the first two panels. 
	The third panel shows the extrapolated intensities
	$I_X^{\uparrow}$, as deduced from the measured intensities $I_X^{\mathrm{avg} }$ and $I_X^{\mathrm{cav}}$ with Eq. (1), and from a self-consistent fit using $P_{\uparrow} = 47 \pm 3 \% = \frac{P_c}{2}$.
	The same extrapolation process is performed for $X$ = $D$, $A$, $R$ and $L$ ( See Supplemental section 11).
	
	The extrapolated behaviour of $\vec{S}_{\uparrow}$ is plotted in the Poincaré sphere as a function of the detuning $\omega_{\mathrm{laser}} - \omega_{\mathrm{QD}}^{\uparrow}$, as shown in Fig. 3b where the left and right spheres represent two views of the same trajectory at B=$1.7\, \mathrm{T}$. 
	$\vec{S}_{\uparrow}$ is experiencing a giant rotation as shown in the first view.
	The second view confirms that $\vec{S}_{\uparrow}$ starts from the polarisation $\ket{V}$, in the off-resonance case, and rotates close to $\ket{H}$ on resonance with the $\ket{\uparrow}$ - $\ket{\uparrow \downarrow \Uparrow}$ transition. 
	Here, the deterministic giant polarisation rotation almost fully reverses the state of the reflected photon, conditioned by the $\ket{\uparrow}$ state of the QD spin.

	The output polarisation purity, described by the norm of the Stokes vector $\vec{S}_{\uparrow}$, is slightly deteriorated on resonance with the QD, as the corresponding points are not at the surface of the sphere. This effect is due to the environmental noise, and in particular the hyperfine interaction between the electron and the surrounding nuclei, leading to fluctuations of $\omega_{\mathrm{QD}}^{\uparrow}$  (See Supplemental sections 6 and 7).
	The behaviour of $\vec{S}_{\uparrow}$ given by the extrapolation is well predicted by our numerical simulation, with the exact same parameters as those used in Fig. 2. 
	In Fig. 3b, we note that some points display an unphysical purity above unity.
	This would not arise with other estimation techniques such as maximum likelihood \cite{Kwiat2004}, yet we chose here the most direct method, displaying $\vec{S}_{\uparrow}$ from the values of its three coordinates $s_{HV}$, $s_{DA}$ and $s_{RL}$, to visualise the result of the extrapolation technique.
	The comparison between the extrapolated data and the numerical simulation confirms the viability of the polarisation tomography in spite of the experimental errors, which include detection noise and nonlinearity, as well as polarisation basis miscalibration. In the results of Fig. 3b, in particular, such errors are amplified by the extrapolation process. \\
	
	\begin{figure}[h!]
		\centering
		\includegraphics[scale=0.22]{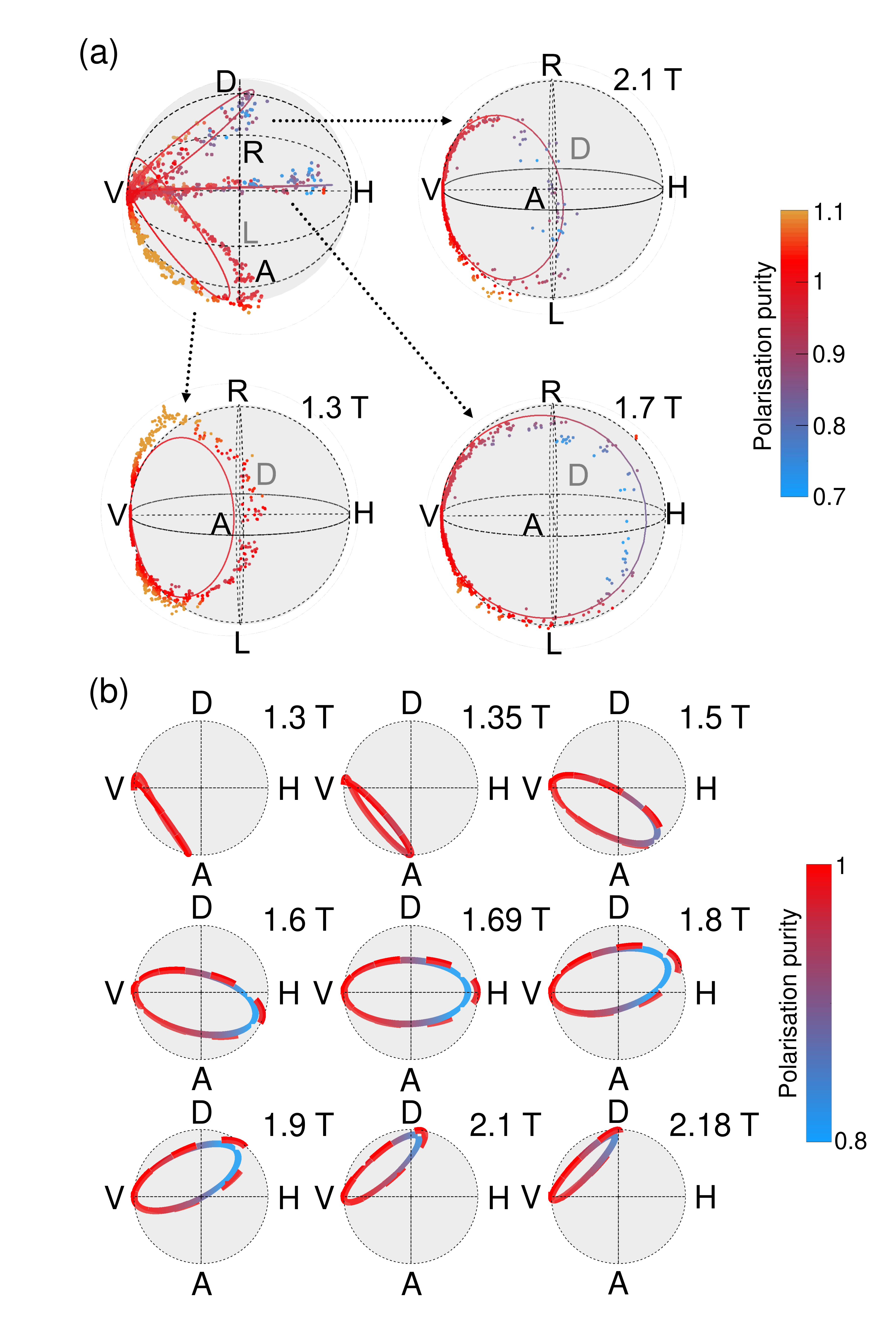}
		\caption{
			\textbf{Generating different output states in the Poincaré sphere} - 
			\textbf{(a)} 
			Extrapolated Stokes vector $\vec{S_{\uparrow}}$, for different magnetic fields (circles: extrapolated from experimental data with $P_c=94\%$; lines: numerical simulations).  In this figure, the trajectories correspond to scan of detunings $\omega_{\mathrm{laser}} - \omega_{\mathrm{QD}}^{\uparrow}$ similarly to Fig. 2c, but the colorscale indicates the polarisation purity. 
			\textbf{(b)} 
			Numerical simulation of the reflected Stokes vector $\vec{S_{\uparrow}}$, viewed from the top of the Poincaré sphere for different magnetic fields without and with environmental noise (resp. solid line and dashed lines).
		}
	\end{figure}
	
	Thus far, the extrapolation process at $1.7\, \mathrm{T}$ already demonstrated the ability to produce different outputs in the Poincaré sphere by controlling the incoming photon energy $\omega_{\mathrm{laser}}$.
	In addition, the trajectory of $\vec{S}_{\uparrow}$ in the Poincaré sphere is also controlled by the value of $\omega_{\mathrm{QD}}^{\uparrow}$, experimentally tunable with the applied magnetic field. 
	The results of the extrapolation processes are displayed in the Poincaré sphere, in Fig. 4a, for the three magnetic fields previously explored.
	The top-left panel aggregates the trajectories of $\vec{S}_{\uparrow}$ for all magnetic fields, viewed from the top of the sphere.
	Each other panel isolates $\vec{S}_{\uparrow}$ for a single magnetic field, under a second viewing angle. 
	In these panels, though the points correspond to different laser-QD detunings, each point is color-mapped with respect to its corresponding polarisation purity.
	Here again, the non-physical points with a purity above unity are explained by the experimental uncertainties on the measurement of $I_X^{\mathrm{avg} }$ and $I_X^{\mathrm{cav}}$, from which $I_X^{\uparrow}$ is deduced. 
	This is especially visible at $1.3\, \mathrm{T}$, where $\omega_{\mathrm{QD}}^{\uparrow}$ is close to $\omega_{\mathrm{cav}, V}$, leading to low reflectivities, and a higher sensitivity to detector noise and residual cavity-induced polarisation rotations. 
	This discrepancy between the theoretical model and the experimental data is amplified by the extrapolation process, limiting its accuracy. A relatively good agreement is still found between the extrapolated trajectories of $\vec{S}_{\uparrow}$ and the numerical simulations.
	
	
	\textbf{Discussion.}
	These experimental results indicate the possibility to generate an arbitrary state, through a proper setting of the experimental parameters. 
	\LL{This is indeed possible because the rotated polarisation state, ideally of the form $\ket{\psi_{\uparrow}}	\propto r_R^{\uparrow}\ket{R} - r_L^{\uparrow}\ket{L}$ in absence of environmental noise, can be entirely controlled by varying the reflection coefficients $r_R^{\uparrow}$ and $r_L^{\uparrow}$. 
		In particular, the phase (resp. amplitude) difference between $r_R^{\uparrow}$ and $r_L^{\uparrow}$ 
		controls the rotation in longitude (resp. latitude) of the Stokes vector $\vec{S}_{\uparrow}$, in the Poincaré sphere. 
		In practice, these two rotations depend on two experimentally-controlled parameters, i.e. $\omega_{\mathrm{laser}}$ and $\omega_{\mathrm{{QD}}}^{\uparrow}$, that can be independently varied, in the vicinity of the two cavity mode energies $\omega_{\mathrm{cav,}H}$ and $\omega_{\mathrm{cav,}V}$. 
		This is how the magnetic fields discussed above have been set, in order to reach, at a proper laser-QD detuning, an output polarisation close to $H$ for $1.7\, \mathrm{T}$ ($\pi$ phase shift induced between $r_R^{\uparrow}$ and $r_L^{\uparrow}$), close to $A$ for $1.3\, \mathrm{T}$ ($\frac{\pi}{2}$ phase shift), and close to $D$ for $2.1\, \mathrm{T}$ ($-\frac{\pi}{2}$ phase shift). }
	
	To illustrate this, we show in Fig. 4b numerical simulations for the trajectory of $\vec{S}_{\uparrow}$ in the Poincaré sphere, with environmental noise ($\sigma = 0.5 ~\mu eV$, solid lines) and without (dashed lines), for other magnetic field amplitudes. 
	For each trajectory, the simulated points are obtained for different detunings between the QD and the laser, yet here again the colorscale represents the polarisation purity.
	In the absence of noise, the polarisation of the reflected photons can reach any state at the surface of the Poincaré sphere, with a unity polarisation purity. 
	In the presence of environmental noise, the expected polarisation is slightly degraded down to 0.81  when reaching a $\pi$ phase shift at 1.69T, i.e. when $\omega_{\mathrm{QD}}^{\uparrow} = \omega_{\mathrm{cav},V}+51.4\,\mu eV$. 
	This is directly related to the averaging of the Stokes vector over various orientations, induced by the fluctuations of $\omega_{\mathrm{{QD}}}^{\uparrow}$.
	
	We note, however, that this depolarisation at $1.69~$T still allows reaching a fidelity of $90.5\%$ (see Methods) with respect to the corresponding ideal state $\ket{H}$ for the reflected photons. 
	Also, a good polarisation purity of $98\%$  is found at 1.35T ($\omega_{\mathrm{QD}}^{\uparrow} = \omega_{\mathrm{cav},V}+14.1\,\mu eV$), in the $ \frac{\pi}{2}$ phase shift configuration: this translates into a fidelity of $99\%$ with respect to the corresponding desired state $\ket{A}$. 
	This is especially promising for reaching perfect spin-polarisation mapping, i.e. $\braket{\Psi_{\uparrow}|\Psi_{\downarrow}}$ =0. If the same QD-cavity detuning had been obtained at zero magnetic field, i.e. with $\omega_{\mathrm{{QD}}}^{\downarrow}=\omega_{\mathrm{{QD}}}^{\uparrow}=\omega_{\mathrm{cav},V}+14.1\,\mu eV$, the spin states $\ket{\uparrow}$ and $\ket{\downarrow}$ would almost have been mapped to the opposite polarisation states  $\ket{\Psi_{\uparrow}}=\ket{A}$ and $\ket{\Psi_{\downarrow}}=\ket{D}$. Being able to operate spin-photon interfaces at zero or low magnetic field is particularly interesting to be compatible with other quantum protocols for the generation of photonic cluster states \cite{Lindner2009}. 
	
	\LL{The giant and stable rotations achieved here had never been observed on pillar-based devices, due to limited output coupling efficiencies (lower than $50\%$) \cite{Arnold2015} and/or due to the impact of environmental noise  \cite{Arnold2015,Androvi2019,Wells2019}. 
		To overcome the effect of slow charge-induced noise \cite{Prechtel2016}, i.e. spectral wandering below the $50\,kHz$ range, a post-selection approach has been developped \cite{Androvi2019}. 
		This technique selects a small portion of the experimental data, and would fail to take into account the fast spin-induced noise \cite{Prechtel2016} caused by the hyperfine interaction with surrounding nuclear spins \cite{Urbaszek2013}. 
		Though being imperfect, the cavity-QED device presented here actually meets all the conditions to generate most polarisation states. This is due to the electrical control stabilizing the charge environment \cite{DeSantis2017}, and to the Purcell enhancement of the QD-transition homogeneous linewidth, above the expected hyperfine-induced fluctuations of the Zeeman splitting. 
		Improving the output coupling efficiency, while maximizing the device cooperativity, would allow achieving purer polarisation states, together with higher total reflectivities, i.e. larger probabilities for a spin-photon gate to successfully reflect the incoming photons. 
		In addition, recent techniques have appeared \cite{Gangloff2019}, allowing drastic reductions of the hyperfine-induced fluctuations, that could be applied to develop truly optimal spin-photon interfaces.}
	
	As detailed in the Supplemental section 8, the possibility of fully exploring the Poincaré sphere is a direct consequence of the moderate birefringence, allowing to enhance the light-matter interaction both for the $H$ and $V$ polarisations. This control offers the possibility of implementing perfect  spin-polarisation mapping at any magnetic fields, by maximizing the distance between $\ket{\Psi_{\uparrow}}$ and $\ket{\Psi_{\downarrow}}$ in the Poincaré sphere, until the  desired condition $\braket{\Psi_{\uparrow} | \Psi_{\downarrow}}=0$ is reached.
	This strongly encourages pursuing the efforts towards new cavity-QED devices with low birefrigence, also including bullseye \cite{Singh2022} and open Fabry-Perot cavities \cite{Antoniadis2022}, or carefully-engineered photonic crystals \cite{Luxmoore2013}. With highly-birefringent cavities \cite{Tiecke2014, Kim2013,Nguyen2019, Javadi2018}, only a portion of the Poincaré sphere can be reached.
	This may not prevent engineering a perfect spin-polarisation mapping, at high magnetic fields and with carefully-chosen parameters (see Supplemental section 9). Yet, generally, the corresponding input and output states will both have non-trivial longitudes and latitudes in the Poincaré sphere. This renders all the more necessary to include the  tomography approach in the various range of experiments involving polarisation-encoded interfaces, to allow adapting the experimental protocols to device imperfections.

	\section*{Methods}
	
	\textbf{Device fabrication.} 
	The sample is grown by molecular beam epitaxy and consists of a $\lambda$-GaAs cavity, formed by two distributed Bragg reflectors, embedding an annealed InGaAs QD \cite{Somaschi2016,DeSantis2017}. 
	The Bragg mirrors are made by alternating layers of GaAs and $\mathrm{Al}_{0.9}\mathrm{Ga}_{0.1}\mathrm{As}$, with 20 (30) pairs for the top (bottom) mirror. 
	To electrically contact the structure, the bottom mirror has a gradual n-doping profile while the top mirror has a p-doping profile. 
	The deterministic spatial and spectral matching between the micropillar cavity and a single QD is achieved by in-situ lithography \cite{Nowak2014}. 
	Each micropillar is connected through four ridges to a large circular frame attached to a gold plated mesa, enabling the electrical control. 
	Applying a bias in the [$-1$~V$;-0.7$~V] range stabilises the electrical environment of the QD, while ensuring a maximal probability for a single electron to occupy it. On the other hand, applying a $0$~V bias allows removing charges from the QD, and thus fully populate state $\ket{\emptyset}$. The micropillar presents a small ellipticity leading to the lift of degeneracy of the cavity modes, which are split in two partially-overlapping modes corresponding to $H$ and $V$ polarisations (See Fig.1b and Supplemental section 4).\\ 
	
	\noindent
	\textbf{Resonant excitation experiments.}
	A tunable continuous wave laser, in the linear low power regime ($P_{\mathrm{in}}  = 4\,pW$), is injected in the QD-micropillar device, placed in a liquid Helium cryostat at $4\,K$. The application of a longitudinal magnetic field is used to vary the $\ket{\uparrow} - \ket{\uparrow \downarrow \Uparrow}$ transition between 925.05$\,$nm and 925.20$\,$nm. Such wavelengths allow obtaining large polarisation rotations especially since they are positioned in-between the two cavity mode resonances at 925.156$\,$nm ($V$-polarised mode) and 925.056$\,$nm ($H$-polarised mode). \\
	
	\noindent
	\textbf{Polarisation characterisation.} 
	The polarisation tomography setup successively measures the average intensities $I_X^{\mathrm{avg}}$ in the six different polarisation bases $X$ = $H$, $V$, $D$, $A$, $R$ and $L$. 
	The Stokes parameters $s_{X\bar{X}}$ are then deduced as
	$s_{X\bar{X}}^{\mathrm{avg}} = (I_X^{\mathrm{avg}} - I_{\bar{X}}^{\mathrm{avg}})/(I_X^{\mathrm{avg}} + I_{\bar{X}}^{\mathrm{avg} })$,
	with $X\bar{X}$ = $HV$, $DA$, $RL$.
	A given polarisation in the Poincaré sphere is characterised by its Stokes \LL{vector with components} ($s_{HV}$, $s_{DA}$, $s_{RL}$), and its polarisation purity by $\rho = \sqrt{s_{HV}^2 + s_{DA}^2 + s_{RL}^2}$. \LL{The fidelity of a polarisation state to an ideal pure state can be expressed in terms of the actual Stokes vector $\vec{S}$ and the ideal target one $\vec{S}_\mathrm{target}$, through $F=\frac{1}{2} (1+\vec{S}.\vec{S}_\mathrm{target})$.}

	\section*{References}
	

\begin{thebibliography}{51}%
		\makeatletter
		\providecommand \@ifxundefined [1]{%
			\@ifx{#1\undefined}
		}%
		\providecommand \@ifnum [1]{%
			\ifnum #1\expandafter \@firstoftwo
			\else \expandafter \@secondoftwo
			\fi
		}%
		\providecommand \@ifx [1]{%
			\ifx #1\expandafter \@firstoftwo
			\else \expandafter \@secondoftwo
			\fi
		}%
		\providecommand \natexlab [1]{#1}%
		\providecommand \enquote  [1]{``#1''}%
		\providecommand \bibnamefont  [1]{#1}%
		\providecommand \bibfnamefont [1]{#1}%
		\providecommand \citenamefont [1]{#1}%
		\providecommand \href@noop [0]{\@secondoftwo}%
		\providecommand \href [0]{\begingroup \@sanitize@url \@href}%
		\providecommand \@href[1]{\@@startlink{#1}\@@href}%
		\providecommand \@@href[1]{\endgroup#1\@@endlink}%
		\providecommand \@sanitize@url [0]{\catcode `\\12\catcode `\$12\catcode
			`\&12\catcode `\#12\catcode `\^12\catcode `\_12\catcode `\%12\relax}%
		\providecommand \@@startlink[1]{}%
		\providecommand \@@endlink[0]{}%
		\providecommand \url  [0]{\begingroup\@sanitize@url \@url }%
		\providecommand \@url [1]{\endgroup\@href {#1}{\urlprefix }}%
		\providecommand \urlprefix  [0]{URL }%
		\providecommand \Eprint [0]{\href }%
		\providecommand \doibase [0]{https://doi.org/}%
		\providecommand \selectlanguage [0]{\@gobble}%
		\providecommand \bibinfo  [0]{\@secondoftwo}%
		\providecommand \bibfield  [0]{\@secondoftwo}%
		\providecommand \translation [1]{[#1]}%
		\providecommand \BibitemOpen [0]{}%
		\providecommand \bibitemStop [0]{}%
		\providecommand \bibitemNoStop [0]{.\EOS\space}%
		\providecommand \EOS [0]{\spacefactor3000\relax}%
		\providecommand \BibitemShut  [1]{\csname bibitem#1\endcsname}%
		\let\auto@bib@innerbib\@empty
		\bibitem [{\citenamefont {Cirac}\ \emph {et~al.}(1997)\citenamefont {Cirac},
			\citenamefont {Zoller}, \citenamefont {Kimble},\ and\ \citenamefont
			{Mabuchi}}]{Cirac1997}%
		\BibitemOpen
		\bibfield  {author} {\bibinfo {author} {\bibfnamefont {J.~I.}\ \bibnamefont
				{Cirac}}, \bibinfo {author} {\bibfnamefont {P.}~\bibnamefont {Zoller}},
			\bibinfo {author} {\bibfnamefont {H.~J.}\ \bibnamefont {Kimble}},\ and\
			\bibinfo {author} {\bibfnamefont {H.}~\bibnamefont {Mabuchi}},\ }\bibfield
		{title} {\bibinfo {title} {Quantum state transfer and entanglement
				distribution among distant nodes in a quantum network},\ }\href@noop {}
		{\bibfield  {journal} {\bibinfo  {journal} {Phys. Rev. Lett.}\ }\textbf
			{\bibinfo {volume} {78}},\ \bibinfo {pages} {3221} (\bibinfo {year}
			{1997})}\BibitemShut {NoStop}%
		\bibitem [{\citenamefont {Borregaard}\ \emph {et~al.}(2020)\citenamefont
			{Borregaard}, \citenamefont {Pichler}, \citenamefont {Schröder},
			\citenamefont {Lukin}, \citenamefont {Lodahl},\ and\ \citenamefont
			{Sørensen}}]{Borregaard2020}%
		\BibitemOpen
		\bibfield  {author} {\bibinfo {author} {\bibfnamefont {J.}~\bibnamefont
				{Borregaard}}, \bibinfo {author} {\bibfnamefont {H.}~\bibnamefont {Pichler}},
			\bibinfo {author} {\bibfnamefont {T.}~\bibnamefont {Schröder}}, \bibinfo
			{author} {\bibfnamefont {M.~D.}\ \bibnamefont {Lukin}}, \bibinfo {author}
			{\bibfnamefont {P.}~\bibnamefont {Lodahl}},\ and\ \bibinfo {author}
			{\bibfnamefont {A.~S.}\ \bibnamefont {Sørensen}},\ }\bibfield  {title}
		{\bibinfo {title} {{One-Way Quantum Repeater Based on Near-Deterministic
					Photon-Emitter Interfaces}},\ }\href@noop {} {\bibfield  {journal} {\bibinfo
				{journal} {{Phys. Rev. X }}\ }\textbf {\bibinfo {volume} {10}},\ \bibinfo
			{pages} {021071} (\bibinfo {year} {2020})}\BibitemShut {NoStop}%
		\bibitem [{\citenamefont {Pichler}\ \emph {et~al.}(2017)\citenamefont
			{Pichler}, \citenamefont {Choi}, \citenamefont {Zoller},\ and\ \citenamefont
			{Lukin}}]{Pichler2017}%
		\BibitemOpen
		\bibfield  {author} {\bibinfo {author} {\bibfnamefont {H.}~\bibnamefont
				{Pichler}}, \bibinfo {author} {\bibfnamefont {S.}~\bibnamefont {Choi}},
			\bibinfo {author} {\bibfnamefont {P.}~\bibnamefont {Zoller}},\ and\ \bibinfo
			{author} {\bibfnamefont {M.~D.}\ \bibnamefont {Lukin}},\ }\bibfield  {title}
		{\bibinfo {title} {{Universal photonic quantum computation via time-delayed
					feedback}},\ }\href@noop {} {\bibfield  {journal} {\bibinfo  {journal}
				{{Proceedings of the National Academy of Science}}\ }\textbf {\bibinfo
				{volume} {114}},\ \bibinfo {pages} {11362} (\bibinfo {year}
			{2017})}\BibitemShut {NoStop}%
		\bibitem [{\citenamefont {Zhan}\ and\ \citenamefont {Sun}(2020)}]{Sun2020}%
		\BibitemOpen
		\bibfield  {author} {\bibinfo {author} {\bibfnamefont {Y.}~\bibnamefont
				{Zhan}}\ and\ \bibinfo {author} {\bibfnamefont {S.}~\bibnamefont {Sun}},\
		}\bibfield  {title} {\bibinfo {title} {{Deterministic Generation of
					Loss-Tolerant Photonic Cluster States with a Single Quantum Emitter}},\
		}\href@noop {} {\bibfield  {journal} {\bibinfo  {journal} {{Phys. Rev. Lett.
				}}\ }\textbf {\bibinfo {volume} {125}},\ \bibinfo {pages} {223601} (\bibinfo
			{year} {2020})}\BibitemShut {NoStop}%
		\bibitem [{\citenamefont {Shi}\ and\ \citenamefont {Waks}(2021)}]{Shi2021}%
		\BibitemOpen
		\bibfield  {author} {\bibinfo {author} {\bibfnamefont {Y.}~\bibnamefont
				{Shi}}\ and\ \bibinfo {author} {\bibfnamefont {E.}~\bibnamefont {Waks}},\
		}\bibfield  {title} {\bibinfo {title} {{Deterministic generation of
					multidimensional photonic cluster states using time-delay feedback}},\
		}\href@noop {} {\bibfield  {journal} {\bibinfo  {journal} {{Phys. Rev. A}}\
			}\textbf {\bibinfo {volume} {104}},\ \bibinfo {pages} {013703} (\bibinfo
			{year} {2021})}\BibitemShut {NoStop}%
		\bibitem [{\citenamefont {Reiserer}\ \emph {et~al.}(2014)\citenamefont
			{Reiserer}, \citenamefont {Kalb}, \citenamefont {Rempe},\ and\ \citenamefont
			{Ritter}}]{Reiserer2014}%
		\BibitemOpen
		\bibfield  {author} {\bibinfo {author} {\bibfnamefont {A.}~\bibnamefont
				{Reiserer}}, \bibinfo {author} {\bibfnamefont {N.}~\bibnamefont {Kalb}},
			\bibinfo {author} {\bibfnamefont {G.}~\bibnamefont {Rempe}},\ and\ \bibinfo
			{author} {\bibfnamefont {S.}~\bibnamefont {Ritter}},\ }\bibfield  {title}
		{\bibinfo {title} {{A quantum gate between a flying optical photon and a
					single trapped atom}},\ }\href@noop {} {\bibfield  {journal} {\bibinfo
				{journal} {{Nature}}\ }\textbf {\bibinfo {volume} {508}},\ \bibinfo {pages}
			{237} (\bibinfo {year} {2014})}\BibitemShut {NoStop}%
		\bibitem [{\citenamefont {Tiecke}\ \emph {et~al.}(2014)\citenamefont {Tiecke},
			\citenamefont {Thompson}, \citenamefont {de~Leon}, \citenamefont {Liu},
			\citenamefont {Vuletić},\ and\ \citenamefont {Lukin}}]{Tiecke2014}%
		\BibitemOpen
		\bibfield  {author} {\bibinfo {author} {\bibfnamefont {T.~G.}\ \bibnamefont
				{Tiecke}}, \bibinfo {author} {\bibfnamefont {J.~D.}\ \bibnamefont
				{Thompson}}, \bibinfo {author} {\bibfnamefont {N.~P.}\ \bibnamefont
				{de~Leon}}, \bibinfo {author} {\bibfnamefont {L.~R.}\ \bibnamefont {Liu}},
			\bibinfo {author} {\bibfnamefont {V.}~\bibnamefont {Vuletić}},\ and\
			\bibinfo {author} {\bibfnamefont {M.~D.}\ \bibnamefont {Lukin}},\ }\bibfield
		{title} {\bibinfo {title} {{Nanophotonic quantum phase switch with a single
					atom}},\ }\href@noop {} {\bibfield  {journal} {\bibinfo  {journal}
				{{Nature}}\ }\textbf {\bibinfo {volume} {508}},\ \bibinfo {pages} {241}
			(\bibinfo {year} {2014})}\BibitemShut {NoStop}%
		\bibitem [{\citenamefont {Shomroni}\ \emph {et~al.}(2014)\citenamefont
			{Shomroni}, \citenamefont {Rosenblum}, \citenamefont {Lovsky}, \citenamefont
			{Bechler}, \citenamefont {Guendelman},\ and\ \citenamefont
			{Dayan}}]{Shomroni2014}%
		\BibitemOpen
		\bibfield  {author} {\bibinfo {author} {\bibfnamefont {I.}~\bibnamefont
				{Shomroni}}, \bibinfo {author} {\bibfnamefont {S.}~\bibnamefont {Rosenblum}},
			\bibinfo {author} {\bibfnamefont {Y.}~\bibnamefont {Lovsky}}, \bibinfo
			{author} {\bibfnamefont {O.}~\bibnamefont {Bechler}}, \bibinfo {author}
			{\bibfnamefont {G.}~\bibnamefont {Guendelman}},\ and\ \bibinfo {author}
			{\bibfnamefont {B.}~\bibnamefont {Dayan}},\ }\bibfield  {title} {\bibinfo
			{title} {{All-optical routing of single photons by a one-atom switch
					controlled by a single photon}},\ }\href@noop {} {\bibfield  {journal}
			{\bibinfo  {journal} {{Science}}\ }\textbf {\bibinfo {volume} {345}},\
			\bibinfo {pages} {903} (\bibinfo {year} {2014})}\BibitemShut {NoStop}%
		\bibitem [{\citenamefont {Bechler}\ \emph {et~al.}(2018)\citenamefont
			{Bechler}, \citenamefont {Borne}, \citenamefont {Rosenblum}, \citenamefont
			{Guendelman}, \citenamefont {Mor}, \citenamefont {Netser}, \citenamefont
			{Ohana}, \citenamefont {Aqua}, \citenamefont {Drucker}, \citenamefont
			{Finkelstein}, \citenamefont {Lovsky}, \citenamefont {Bruch}, \citenamefont
			{Gurovich}, \citenamefont {Shafir},\ and\ \citenamefont
			{Dayan}}]{Bechler2018}%
		\BibitemOpen
		\bibfield  {author} {\bibinfo {author} {\bibfnamefont {O.}~\bibnamefont
				{Bechler}}, \bibinfo {author} {\bibfnamefont {A.}~\bibnamefont {Borne}},
			\bibinfo {author} {\bibfnamefont {S.}~\bibnamefont {Rosenblum}}, \bibinfo
			{author} {\bibfnamefont {G.}~\bibnamefont {Guendelman}}, \bibinfo {author}
			{\bibfnamefont {O.~E.}\ \bibnamefont {Mor}}, \bibinfo {author} {\bibfnamefont
				{M.}~\bibnamefont {Netser}}, \bibinfo {author} {\bibfnamefont
				{T.}~\bibnamefont {Ohana}}, \bibinfo {author} {\bibfnamefont
				{Z.}~\bibnamefont {Aqua}}, \bibinfo {author} {\bibfnamefont {N.}~\bibnamefont
				{Drucker}}, \bibinfo {author} {\bibfnamefont {R.}~\bibnamefont
				{Finkelstein}}, \bibinfo {author} {\bibfnamefont {Y.}~\bibnamefont {Lovsky}},
			\bibinfo {author} {\bibfnamefont {R.}~\bibnamefont {Bruch}}, \bibinfo
			{author} {\bibfnamefont {D.}~\bibnamefont {Gurovich}}, \bibinfo {author}
			{\bibfnamefont {E.}~\bibnamefont {Shafir}},\ and\ \bibinfo {author}
			{\bibfnamefont {B.}~\bibnamefont {Dayan}},\ }\bibfield  {title} {\bibinfo
			{title} {{A passive photon-atom qubit swap operation}},\ }\href@noop {}
		{\bibfield  {journal} {\bibinfo  {journal} {{Nature Physics}}\ }\textbf
			{\bibinfo {volume} {14}},\ \bibinfo {pages} {996} (\bibinfo {year}
			{2018})}\BibitemShut {NoStop}%
		\bibitem [{\citenamefont {Kim}\ \emph {et~al.}(2013)\citenamefont {Kim},
			\citenamefont {Bose}, \citenamefont {Shen}, \citenamefont {Solomon},\ and\
			\citenamefont {Waks}}]{Kim2013}%
		\BibitemOpen
		\bibfield  {author} {\bibinfo {author} {\bibfnamefont {H.}~\bibnamefont
				{Kim}}, \bibinfo {author} {\bibfnamefont {R.}~\bibnamefont {Bose}}, \bibinfo
			{author} {\bibfnamefont {T.~C.}\ \bibnamefont {Shen}}, \bibinfo {author}
			{\bibfnamefont {G.~S.}\ \bibnamefont {Solomon}},\ and\ \bibinfo {author}
			{\bibfnamefont {E.}~\bibnamefont {Waks}},\ }\bibfield  {title} {\bibinfo
			{title} {{A quantum logic gate between a solid-state quantum bit and a
					photon}},\ }\href@noop {} {\bibfield  {journal} {\bibinfo  {journal} {{Nature
						Photon.}}\ }\textbf {\bibinfo {volume} {7}},\ \bibinfo {pages} {373}
			(\bibinfo {year} {2013})}\BibitemShut {NoStop}%
		\bibitem [{\citenamefont {Sun}\ \emph {et~al.}(2016)\citenamefont {Sun},
			\citenamefont {Kim}, \citenamefont {Solomon},\ and\ \citenamefont
			{Waks}}]{Sun2016}%
		\BibitemOpen
		\bibfield  {author} {\bibinfo {author} {\bibfnamefont {S.}~\bibnamefont
				{Sun}}, \bibinfo {author} {\bibfnamefont {H.}~\bibnamefont {Kim}}, \bibinfo
			{author} {\bibfnamefont {G.~S.}\ \bibnamefont {Solomon}},\ and\ \bibinfo
			{author} {\bibfnamefont {E.}~\bibnamefont {Waks}},\ }\bibfield  {title}
		{\bibinfo {title} {{A quantum phase switch between a single solid-state spin
					and a photon}},\ }\href@noop {} {\bibfield  {journal} {\bibinfo  {journal}
				{{Nature Nano. }}\ }\textbf {\bibinfo {volume} {11}},\ \bibinfo {pages} {539}
			(\bibinfo {year} {2016})}\BibitemShut {NoStop}%
		\bibitem [{\citenamefont {Nguyen}\ \emph {et~al.}(2019)\citenamefont {Nguyen},
			\citenamefont {Sukachev}, \citenamefont {Bhaskar}, \citenamefont {Machielse},
			\citenamefont {Levonian}, \citenamefont {Knall}, \citenamefont {Stroganov},
			\citenamefont {Chia}, \citenamefont {Burek}, \citenamefont {Riedinger},
			\citenamefont {Park}, \citenamefont {Lončar},\ and\ \citenamefont
			{Lukin}}]{Nguyen2019}%
		\BibitemOpen
		\bibfield  {author} {\bibinfo {author} {\bibfnamefont {C.~T.}\ \bibnamefont
				{Nguyen}}, \bibinfo {author} {\bibfnamefont {D.~D.}\ \bibnamefont
				{Sukachev}}, \bibinfo {author} {\bibfnamefont {M.~K.}\ \bibnamefont
				{Bhaskar}}, \bibinfo {author} {\bibfnamefont {B.}~\bibnamefont {Machielse}},
			\bibinfo {author} {\bibfnamefont {D.~S.}\ \bibnamefont {Levonian}}, \bibinfo
			{author} {\bibfnamefont {E.~N.}\ \bibnamefont {Knall}}, \bibinfo {author}
			{\bibfnamefont {P.}~\bibnamefont {Stroganov}}, \bibinfo {author}
			{\bibfnamefont {C.}~\bibnamefont {Chia}}, \bibinfo {author} {\bibfnamefont
				{M.~J.}\ \bibnamefont {Burek}}, \bibinfo {author} {\bibfnamefont
				{R.}~\bibnamefont {Riedinger}}, \bibinfo {author} {\bibfnamefont
				{H.}~\bibnamefont {Park}}, \bibinfo {author} {\bibfnamefont {M.}~\bibnamefont
				{Lončar}},\ and\ \bibinfo {author} {\bibfnamefont {M.~D.}\ \bibnamefont
				{Lukin}},\ }\bibfield  {title} {\bibinfo {title} {{An integrated nanophotonic
					quantum register based on silicon-vacancy spins in diamond}},\ }\href@noop {}
		{\bibfield  {journal} {\bibinfo  {journal} {{Phys. Rev. B}}\ }\textbf
			{\bibinfo {volume} {100}},\ \bibinfo {pages} {165428} (\bibinfo {year}
			{2019})}\BibitemShut {NoStop}%
		\bibitem [{\citenamefont {Chan}\ \emph {et~al.}(2022)\citenamefont {Chan},
			\citenamefont {Tiranov}, \citenamefont {Appel}, \citenamefont {Wang},
			\citenamefont {Midolo}, \citenamefont {Scholz}, \citenamefont {Wieck},
			\citenamefont {Ludwig}, \citenamefont {Sørensen},\ and\ \citenamefont
			{Lodahl}}]{Chan2022}%
		\BibitemOpen
		\bibfield  {author} {\bibinfo {author} {\bibfnamefont {M.~L.}\ \bibnamefont
				{Chan}}, \bibinfo {author} {\bibfnamefont {A.}~\bibnamefont {Tiranov}},
			\bibinfo {author} {\bibfnamefont {M.~H.}\ \bibnamefont {Appel}}, \bibinfo
			{author} {\bibfnamefont {Y.}~\bibnamefont {Wang}}, \bibinfo {author}
			{\bibfnamefont {L.}~\bibnamefont {Midolo}}, \bibinfo {author} {\bibfnamefont
				{S.}~\bibnamefont {Scholz}}, \bibinfo {author} {\bibfnamefont {A.~D.}\
				\bibnamefont {Wieck}}, \bibinfo {author} {\bibfnamefont {A.}~\bibnamefont
				{Ludwig}}, \bibinfo {author} {\bibfnamefont {A.~S.}\ \bibnamefont
				{Sørensen}},\ and\ \bibinfo {author} {\bibfnamefont {P.}~\bibnamefont
				{Lodahl}},\ }\bibfield  {title} {\bibinfo {title} {{On-chip spin-photon
					entanglement based on single-photon scattering}},\ }\href@noop {} {\bibfield
			{journal} {\bibinfo  {journal} {{arXiv:2205.12844}}\ } (\bibinfo {year}
			{2022})}\BibitemShut {NoStop}%
		\bibitem [{\citenamefont {Antoniadis}\ \emph {et~al.}(2022)\citenamefont
			{Antoniadis}, \citenamefont {Tomm}, \citenamefont {Jakubczyk}, \citenamefont
			{Schott}, \citenamefont {Valentin}, \citenamefont {Wieck}, \citenamefont
			{Ludwig}, \citenamefont {Warburton},\ and\ \citenamefont
			{Javadi}}]{Antoniadis2022}%
		\BibitemOpen
		\bibfield  {author} {\bibinfo {author} {\bibfnamefont {N.~O.}\ \bibnamefont
				{Antoniadis}}, \bibinfo {author} {\bibfnamefont {N.}~\bibnamefont {Tomm}},
			\bibinfo {author} {\bibfnamefont {T.}~\bibnamefont {Jakubczyk}}, \bibinfo
			{author} {\bibfnamefont {R.}~\bibnamefont {Schott}}, \bibinfo {author}
			{\bibfnamefont {S.~R.}\ \bibnamefont {Valentin}}, \bibinfo {author}
			{\bibfnamefont {A.~D.}\ \bibnamefont {Wieck}}, \bibinfo {author}
			{\bibfnamefont {A.}~\bibnamefont {Ludwig}}, \bibinfo {author} {\bibfnamefont
				{R.~J.}\ \bibnamefont {Warburton}},\ and\ \bibinfo {author} {\bibfnamefont
				{A.}~\bibnamefont {Javadi}},\ }\bibfield  {title} {\bibinfo {title} {{A
					chiral one-dimensional atom using a quantum dot in an open microcavity}},\
		}\href@noop {} {\bibfield  {journal} {\bibinfo  {journal} {{npj Quantum
						Inf}}\ }\textbf {\bibinfo {volume} {8}},\ \bibinfo {pages} {27} (\bibinfo
			{year} {2022})}\BibitemShut {NoStop}%
		\bibitem [{\citenamefont {Reuer}\ \emph {et~al.}(2022)\citenamefont {Reuer},
			\citenamefont {Besse}, \citenamefont {Wernli}, \citenamefont {Magnard},
			\citenamefont {Kurpiers}, \citenamefont {Norris}, \citenamefont {Wallraff},\
			and\ \citenamefont {Eichler}}]{Reuer2022}%
		\BibitemOpen
		\bibfield  {author} {\bibinfo {author} {\bibfnamefont {K.}~\bibnamefont
				{Reuer}}, \bibinfo {author} {\bibfnamefont {J.-C.}\ \bibnamefont {Besse}},
			\bibinfo {author} {\bibfnamefont {L.}~\bibnamefont {Wernli}}, \bibinfo
			{author} {\bibfnamefont {P.}~\bibnamefont {Magnard}}, \bibinfo {author}
			{\bibfnamefont {P.}~\bibnamefont {Kurpiers}}, \bibinfo {author}
			{\bibfnamefont {G.~J.}\ \bibnamefont {Norris}}, \bibinfo {author}
			{\bibfnamefont {A.}~\bibnamefont {Wallraff}},\ and\ \bibinfo {author}
			{\bibfnamefont {C.}~\bibnamefont {Eichler}},\ }\bibfield  {title} {\bibinfo
			{title} {{Realization of a Universal Quantum Gate Set for Itinerant Microwave
					Photons}},\ }\href@noop {} {\bibfield  {journal} {\bibinfo  {journal} {{Phys.
						Rev. X}}\ }\textbf {\bibinfo {volume} {12}},\ \bibinfo {pages} {011008}
			(\bibinfo {year} {2022})}\BibitemShut {NoStop}%
		\bibitem [{\citenamefont {Hacker}\ \emph {et~al.}(2016)\citenamefont {Hacker},
			\citenamefont {Welte}, \citenamefont {Rempe},\ and\ \citenamefont
			{Ritter}}]{Hacker2016}%
		\BibitemOpen
		\bibfield  {author} {\bibinfo {author} {\bibfnamefont {B.}~\bibnamefont
				{Hacker}}, \bibinfo {author} {\bibfnamefont {S.}~\bibnamefont {Welte}},
			\bibinfo {author} {\bibfnamefont {G.}~\bibnamefont {Rempe}},\ and\ \bibinfo
			{author} {\bibfnamefont {S.}~\bibnamefont {Ritter}},\ }\bibfield  {title}
		{\bibinfo {title} {{A photon–photon quantum gate based on a single atom in
					an optical resonator}},\ }\href@noop {} {\bibfield  {journal} {\bibinfo
				{journal} {{Nature}}\ }\textbf {\bibinfo {volume} {536}},\ \bibinfo {pages}
			{193} (\bibinfo {year} {2016})}\BibitemShut {NoStop}%
		\bibitem [{\citenamefont {Stolz}\ \emph {et~al.}(2022)\citenamefont {Stolz},
			\citenamefont {Hegels}, \citenamefont {Winter}, \citenamefont {Röhr},
			\citenamefont {Hsiao}, \citenamefont {Husel}, \citenamefont {Rempe},\ and\
			\citenamefont {Dürr}}]{Stolz2022}%
		\BibitemOpen
		\bibfield  {author} {\bibinfo {author} {\bibfnamefont {T.}~\bibnamefont
				{Stolz}}, \bibinfo {author} {\bibfnamefont {H.}~\bibnamefont {Hegels}},
			\bibinfo {author} {\bibfnamefont {M.}~\bibnamefont {Winter}}, \bibinfo
			{author} {\bibfnamefont {B.}~\bibnamefont {Röhr}}, \bibinfo {author}
			{\bibfnamefont {Y.-F.}\ \bibnamefont {Hsiao}}, \bibinfo {author}
			{\bibfnamefont {L.}~\bibnamefont {Husel}}, \bibinfo {author} {\bibfnamefont
				{G.}~\bibnamefont {Rempe}},\ and\ \bibinfo {author} {\bibfnamefont
				{S.}~\bibnamefont {Dürr}},\ }\bibfield  {title} {\bibinfo {title}
			{{Quantum-Logic Gate between Two Optical Photons with an Average Efficiency
					above 40\%}},\ }\href@noop {} {\bibfield  {journal} {\bibinfo  {journal}
				{{Phys. Rev. X}}\ }\textbf {\bibinfo {volume} {12}},\ \bibinfo {pages}
			{021035} (\bibinfo {year} {2022})}\BibitemShut {NoStop}%
		\bibitem [{\citenamefont {Sun}\ \emph {et~al.}(2018)\citenamefont {Sun},
			\citenamefont {Kim}, \citenamefont {Luo}, \citenamefont {Solomon},\ and\
			\citenamefont {Waks}}]{Sun2018}%
		\BibitemOpen
		\bibfield  {author} {\bibinfo {author} {\bibfnamefont {S.}~\bibnamefont
				{Sun}}, \bibinfo {author} {\bibfnamefont {H.}~\bibnamefont {Kim}}, \bibinfo
			{author} {\bibfnamefont {Z.}~\bibnamefont {Luo}}, \bibinfo {author}
			{\bibfnamefont {G.~S.}\ \bibnamefont {Solomon}},\ and\ \bibinfo {author}
			{\bibfnamefont {E.}~\bibnamefont {Waks}},\ }\bibfield  {title} {\bibinfo
			{title} {{A single-photon switch and transistor enabled by a solid-state
					quantum memory}},\ }\href@noop {} {\bibfield  {journal} {\bibinfo  {journal}
				{{Science}}\ }\textbf {\bibinfo {volume} {361}},\ \bibinfo {pages} {57}
			(\bibinfo {year} {2018})}\BibitemShut {NoStop}%
		\bibitem [{\citenamefont {Wang}\ \emph {et~al.}(2022)\citenamefont {Wang},
			\citenamefont {Bao}, \citenamefont {Li}, \citenamefont {Wu}, \citenamefont
			{Cai}, \citenamefont {Wang}, \citenamefont {Han}, \citenamefont {Wang},
			\citenamefont {Song}, \citenamefont {Sun}, \citenamefont {Zhang},\ and\
			\citenamefont {Duan}}]{Wang2022}%
		\BibitemOpen
		\bibfield  {author} {\bibinfo {author} {\bibfnamefont {Z.}~\bibnamefont
				{Wang}}, \bibinfo {author} {\bibfnamefont {Z.}~\bibnamefont {Bao}}, \bibinfo
			{author} {\bibfnamefont {Y.}~\bibnamefont {Li}}, \bibinfo {author}
			{\bibfnamefont {Y.}~\bibnamefont {Wu}}, \bibinfo {author} {\bibfnamefont
				{W.}~\bibnamefont {Cai}}, \bibinfo {author} {\bibfnamefont {W.}~\bibnamefont
				{Wang}}, \bibinfo {author} {\bibfnamefont {X.}~\bibnamefont {Han}}, \bibinfo
			{author} {\bibfnamefont {J.}~\bibnamefont {Wang}}, \bibinfo {author}
			{\bibfnamefont {Y.}~\bibnamefont {Song}}, \bibinfo {author} {\bibfnamefont
				{L.}~\bibnamefont {Sun}}, \bibinfo {author} {\bibfnamefont {H.}~\bibnamefont
				{Zhang}},\ and\ \bibinfo {author} {\bibfnamefont {L.}~\bibnamefont {Duan}},\
		}\bibfield  {title} {\bibinfo {title} {{An ultra-high gain single-photon
					transistor in the microwave regime}},\ }\href@noop {} {\bibfield  {journal}
			{\bibinfo  {journal} {{Nat Commun}}\ }\textbf {\bibinfo {volume} {13}},\
			\bibinfo {pages} {6104} (\bibinfo {year} {2022})}\BibitemShut {NoStop}%
		\bibitem [{\citenamefont {Bhaskar}\ \emph {et~al.}(2020)\citenamefont
			{Bhaskar}, \citenamefont {Riedinger}, \citenamefont {Machielse},
			\citenamefont {Levonian}, \citenamefont {Nguyen}, \citenamefont {Knall},
			\citenamefont {Park}, \citenamefont {Englund}, \citenamefont {Lončar},
			\citenamefont {Sukachev},\ and\ \citenamefont {Lukin}}]{Bhaskar2020}%
		\BibitemOpen
		\bibfield  {author} {\bibinfo {author} {\bibfnamefont {M.~K.}\ \bibnamefont
				{Bhaskar}}, \bibinfo {author} {\bibfnamefont {R.}~\bibnamefont {Riedinger}},
			\bibinfo {author} {\bibfnamefont {B.}~\bibnamefont {Machielse}}, \bibinfo
			{author} {\bibfnamefont {D.~S.}\ \bibnamefont {Levonian}}, \bibinfo {author}
			{\bibfnamefont {C.~T.}\ \bibnamefont {Nguyen}}, \bibinfo {author}
			{\bibfnamefont {E.~N.}\ \bibnamefont {Knall}}, \bibinfo {author}
			{\bibfnamefont {H.}~\bibnamefont {Park}}, \bibinfo {author} {\bibfnamefont
				{D.}~\bibnamefont {Englund}}, \bibinfo {author} {\bibfnamefont
				{M.}~\bibnamefont {Lončar}}, \bibinfo {author} {\bibfnamefont {D.~D.}\
				\bibnamefont {Sukachev}},\ and\ \bibinfo {author} {\bibfnamefont {M.~D.}\
				\bibnamefont {Lukin}},\ }\bibfield  {title} {\bibinfo {title} {{Experimental
					demonstration of memory-enhanced quantum communication}},\ }\href@noop {}
		{\bibfield  {journal} {\bibinfo  {journal} {{Nature }}\ }\textbf {\bibinfo
				{volume} {580}},\ \bibinfo {pages} {60–64} (\bibinfo {year}
			{2020})}\BibitemShut {NoStop}%
		\bibitem [{\citenamefont {Kono}\ \emph {et~al.}(2018)\citenamefont {Kono},
			\citenamefont {Koshino}, \citenamefont {Tabuchi}, \citenamefont {Noguchi},\
			and\ \citenamefont {Nakamura}}]{Kono2018}%
		\BibitemOpen
		\bibfield  {author} {\bibinfo {author} {\bibfnamefont {S.}~\bibnamefont
				{Kono}}, \bibinfo {author} {\bibfnamefont {K.}~\bibnamefont {Koshino}},
			\bibinfo {author} {\bibfnamefont {Y.}~\bibnamefont {Tabuchi}}, \bibinfo
			{author} {\bibfnamefont {A.}~\bibnamefont {Noguchi}},\ and\ \bibinfo {author}
			{\bibfnamefont {Y.}~\bibnamefont {Nakamura}},\ }\bibfield  {title} {\bibinfo
			{title} {{Quantum non-demolition detection of an itinerant microwave
					photon}},\ }\href@noop {} {\bibfield  {journal} {\bibinfo  {journal} {{Nature
						Phys}}\ }\textbf {\bibinfo {volume} {14}},\ \bibinfo {pages} {546–549}
			(\bibinfo {year} {2018})}\BibitemShut {NoStop}%
		\bibitem [{\citenamefont {Atatüre}\ \emph {et~al.}(2018)\citenamefont
			{Atatüre}, \citenamefont {Englund}, \citenamefont {Vamivakas}, \citenamefont
			{Lee},\ and\ \citenamefont {Wrachtrup}}]{Atature2018}%
		\BibitemOpen
		\bibfield  {author} {\bibinfo {author} {\bibfnamefont {M.}~\bibnamefont
				{Atatüre}}, \bibinfo {author} {\bibfnamefont {D.}~\bibnamefont {Englund}},
			\bibinfo {author} {\bibfnamefont {N.}~\bibnamefont {Vamivakas}}, \bibinfo
			{author} {\bibfnamefont {S.-Y.}\ \bibnamefont {Lee}},\ and\ \bibinfo {author}
			{\bibfnamefont {J.}~\bibnamefont {Wrachtrup}},\ }\bibfield  {title} {\bibinfo
			{title} {{Material platforms for spin-based photonic quantum technologies}},\
		}\href@noop {} {\bibfield  {journal} {\bibinfo  {journal} {{Nat Rev Mater}}\
			}\textbf {\bibinfo {volume} {3}},\ \bibinfo {pages} {38–51} (\bibinfo
			{year} {2018})}\BibitemShut {NoStop}%
		\bibitem [{\citenamefont {Awschalom}\ \emph {et~al.}(2018)\citenamefont
			{Awschalom}, \citenamefont {Hanson}, \citenamefont {Wrachtrup},\ and\
			\citenamefont {Zhou}}]{Awschalom2018}%
		\BibitemOpen
		\bibfield  {author} {\bibinfo {author} {\bibfnamefont {D.~D.}\ \bibnamefont
				{Awschalom}}, \bibinfo {author} {\bibfnamefont {R.}~\bibnamefont {Hanson}},
			\bibinfo {author} {\bibfnamefont {J.}~\bibnamefont {Wrachtrup}},\ and\
			\bibinfo {author} {\bibfnamefont {B.~B.}\ \bibnamefont {Zhou}},\ }\bibfield
		{title} {\bibinfo {title} {{Quantum technologies with optically interfaced
					solid-state spins}},\ }\href@noop {} {\bibfield  {journal} {\bibinfo
				{journal} {{Nature Photon.}}\ }\textbf {\bibinfo {volume} {12}},\ \bibinfo
			{pages} {516–527} (\bibinfo {year} {2018})}\BibitemShut {NoStop}%
		\bibitem [{\citenamefont {Leuenberger}(2006)}]{Leuenberger2006}%
		\BibitemOpen
		\bibfield  {author} {\bibinfo {author} {\bibfnamefont {M.~N.}\ \bibnamefont
				{Leuenberger}},\ }\bibfield  {title} {\bibinfo {title} {{Fault-tolerant
					quantum computing with coded spins using the conditional Faraday rotation in
					quantum dots}},\ }\href@noop {} {\bibfield  {journal} {\bibinfo  {journal}
				{{Phys. Rev. B}}\ }\textbf {\bibinfo {volume} {73}},\ \bibinfo {pages}
			{075312} (\bibinfo {year} {2006})}\BibitemShut {NoStop}%
		\bibitem [{\citenamefont {Hu}\ \emph {et~al.}(2008)\citenamefont {Hu},
			\citenamefont {Young}, \citenamefont {O’Brien}, \citenamefont {Munro},\
			and\ \citenamefont {Rarity}}]{Hu2008}%
		\BibitemOpen
		\bibfield  {author} {\bibinfo {author} {\bibfnamefont {C.~Y.}\ \bibnamefont
				{Hu}}, \bibinfo {author} {\bibfnamefont {A.}~\bibnamefont {Young}}, \bibinfo
			{author} {\bibfnamefont {J.~L.}\ \bibnamefont {O’Brien}}, \bibinfo {author}
			{\bibfnamefont {W.~J.}\ \bibnamefont {Munro}},\ and\ \bibinfo {author}
			{\bibfnamefont {J.~G.}\ \bibnamefont {Rarity}},\ }\bibfield  {title}
		{\bibinfo {title} {{Giant optical Faraday rotation induced by a
					single-electron spin in a quantum dot: Applications to entangling remote
					spins via a single photon}},\ }\href@noop {} {\bibfield  {journal} {\bibinfo
				{journal} {{Phys. Rev. B }}\ }\textbf {\bibinfo {volume} {78}},\ \bibinfo
			{pages} {085307} (\bibinfo {year} {2008})}\BibitemShut {NoStop}%
		\bibitem [{\citenamefont {Lindner}\ and\ \citenamefont
			{Rudolph}(2009)}]{Lindner2009}%
		\BibitemOpen
		\bibfield  {author} {\bibinfo {author} {\bibfnamefont {N.~H.}\ \bibnamefont
				{Lindner}}\ and\ \bibinfo {author} {\bibfnamefont {T.}~\bibnamefont
				{Rudolph}},\ }\bibfield  {title} {\bibinfo {title} {{Proposal for pulsed
					on-demand sources of photonic cluster state strings}},\ }\href@noop {}
		{\bibfield  {journal} {\bibinfo  {journal} {{Phys. Rev. Lett. }}\ }\textbf
			{\bibinfo {volume} {103}},\ \bibinfo {pages} {113602} (\bibinfo {year}
			{2009})}\BibitemShut {NoStop}%
		\bibitem [{\citenamefont {Chen}\ \emph {et~al.}(2021)\citenamefont {Chen},
			\citenamefont {Bersin},\ and\ \citenamefont {Englund}}]{Chen2021}%
		\BibitemOpen
		\bibfield  {author} {\bibinfo {author} {\bibfnamefont {K.~C.}\ \bibnamefont
				{Chen}}, \bibinfo {author} {\bibfnamefont {E.}~\bibnamefont {Bersin}},\ and\
			\bibinfo {author} {\bibfnamefont {D.}~\bibnamefont {Englund}},\ }\bibfield
		{title} {\bibinfo {title} {{A polarization encoded photon-to-spin
					interface}},\ }\href@noop {} {\bibfield  {journal} {\bibinfo  {journal} {{npj
						Quantum Inf}}\ }\textbf {\bibinfo {volume} {7}} (\bibinfo {year}
			{2021})}\BibitemShut {NoStop}%
		\bibitem [{\citenamefont {Duan}\ and\ \citenamefont {Kimble}(2004)}]{Duan2004}%
		\BibitemOpen
		\bibfield  {author} {\bibinfo {author} {\bibfnamefont {L.-M.}\ \bibnamefont
				{Duan}}\ and\ \bibinfo {author} {\bibfnamefont {H.~J.}\ \bibnamefont
				{Kimble}},\ }\bibfield  {title} {\bibinfo {title} {{Scalable Photonic Quantum
					Computation through Cavity-Assisted Interactions}},\ }\href@noop {}
		{\bibfield  {journal} {\bibinfo  {journal} {{Phys. Rev. Lett.}}\ }\textbf
			{\bibinfo {volume} {92}},\ \bibinfo {pages} {127902} (\bibinfo {year}
			{2004})}\BibitemShut {NoStop}%
		\bibitem [{\citenamefont {Rakher}\ \emph {et~al.}(2009)\citenamefont {Rakher},
			\citenamefont {Stoltz}, \citenamefont {Coldren}, \citenamefont {Petroff},\
			and\ \citenamefont {Bouwmeester}}]{Rakher2009}%
		\BibitemOpen
		\bibfield  {author} {\bibinfo {author} {\bibfnamefont {M.~T.}\ \bibnamefont
				{Rakher}}, \bibinfo {author} {\bibfnamefont {N.~G.}\ \bibnamefont {Stoltz}},
			\bibinfo {author} {\bibfnamefont {L.~A.}\ \bibnamefont {Coldren}}, \bibinfo
			{author} {\bibfnamefont {P.~M.}\ \bibnamefont {Petroff}},\ and\ \bibinfo
			{author} {\bibfnamefont {D.}~\bibnamefont {Bouwmeester}},\ }\bibfield
		{title} {\bibinfo {title} {{Externally mode-matched cavity quantum
					electrodynamics with charge-tunable quantum dots}},\ }\href@noop {}
		{\bibfield  {journal} {\bibinfo  {journal} {{Phys. Rev. Lett. }}\ }\textbf
			{\bibinfo {volume} {102}},\ \bibinfo {pages} {097403} (\bibinfo {year}
			{2009})}\BibitemShut {NoStop}%
		\bibitem [{\citenamefont {Arnold}\ \emph {et~al.}(2015)\citenamefont {Arnold},
			\citenamefont {Demory}, \citenamefont {Loo}, \citenamefont {Lemaître},
			\citenamefont {Sagnes}, \citenamefont {Glazov}, \citenamefont {Krebs},
			\citenamefont {Voisin}, \citenamefont {Senellart},\ and\ \citenamefont
			{Lanco}}]{Arnold2015}%
		\BibitemOpen
		\bibfield  {author} {\bibinfo {author} {\bibfnamefont {C.}~\bibnamefont
				{Arnold}}, \bibinfo {author} {\bibfnamefont {J.}~\bibnamefont {Demory}},
			\bibinfo {author} {\bibfnamefont {V.}~\bibnamefont {Loo}}, \bibinfo {author}
			{\bibfnamefont {A.}~\bibnamefont {Lemaître}}, \bibinfo {author}
			{\bibfnamefont {I.}~\bibnamefont {Sagnes}}, \bibinfo {author} {\bibfnamefont
				{M.}~\bibnamefont {Glazov}}, \bibinfo {author} {\bibfnamefont
				{O.}~\bibnamefont {Krebs}}, \bibinfo {author} {\bibfnamefont
				{P.}~\bibnamefont {Voisin}}, \bibinfo {author} {\bibfnamefont
				{P.}~\bibnamefont {Senellart}},\ and\ \bibinfo {author} {\bibfnamefont
				{L.}~\bibnamefont {Lanco}},\ }\bibfield  {title} {\bibinfo {title}
			{{Macroscopic rotation of photon polarization induced by a single spin}},\
		}\href@noop {} {\bibfield  {journal} {\bibinfo  {journal} {{Nature Commun.
				}}\ }\textbf {\bibinfo {volume} {6}},\ \bibinfo {pages} {6236} (\bibinfo
			{year} {2015})}\BibitemShut {NoStop}%
		\bibitem [{\citenamefont {Androvitsaneas}\ \emph {et~al.}(2019)\citenamefont
			{Androvitsaneas}, \citenamefont {Young}, \citenamefont {Lennon},
			\citenamefont {Schneider}, \citenamefont {Maier}, \citenamefont {Hinchliff},
			\citenamefont {Atkinson}, \citenamefont {Harbord}, \citenamefont {Kamp},
			\citenamefont {Höfling}, \citenamefont {Rarity},\ and\ \citenamefont
			{Oulton}}]{Androvi2019}%
		\BibitemOpen
		\bibfield  {author} {\bibinfo {author} {\bibfnamefont {P.}~\bibnamefont
				{Androvitsaneas}}, \bibinfo {author} {\bibfnamefont {A.~B.}\ \bibnamefont
				{Young}}, \bibinfo {author} {\bibfnamefont {J.~M.}\ \bibnamefont {Lennon}},
			\bibinfo {author} {\bibfnamefont {C.}~\bibnamefont {Schneider}}, \bibinfo
			{author} {\bibfnamefont {S.}~\bibnamefont {Maier}}, \bibinfo {author}
			{\bibfnamefont {J.~J.}\ \bibnamefont {Hinchliff}}, \bibinfo {author}
			{\bibfnamefont {G.~S.}\ \bibnamefont {Atkinson}}, \bibinfo {author}
			{\bibfnamefont {E.}~\bibnamefont {Harbord}}, \bibinfo {author} {\bibfnamefont
				{M.}~\bibnamefont {Kamp}}, \bibinfo {author} {\bibfnamefont {S.}~\bibnamefont
				{Höfling}}, \bibinfo {author} {\bibfnamefont {J.~G.}\ \bibnamefont
				{Rarity}},\ and\ \bibinfo {author} {\bibfnamefont {R.}~\bibnamefont
				{Oulton}},\ }\bibfield  {title} {\bibinfo {title} {{Efficient quantum
					photonic phase shift in a low Q-factor regime}},\ }\href@noop {} {\bibfield
			{journal} {\bibinfo  {journal} {{ACS Photonics}}\ }\textbf {\bibinfo {volume}
				{6}},\ \bibinfo {pages} {429} (\bibinfo {year} {2019})}\BibitemShut {NoStop}%
		\bibitem [{\citenamefont {Wells}\ \emph {et~al.}(2019)\citenamefont {Wells},
			\citenamefont {Kalliakos}, \citenamefont {Villa}, \citenamefont {Ellis},
			\citenamefont {Stevenson}, \citenamefont {Bennett}, \citenamefont {Farrer},
			\citenamefont {Ritchie},\ and\ \citenamefont {Shields}}]{Wells2019}%
		\BibitemOpen
		\bibfield  {author} {\bibinfo {author} {\bibfnamefont {L.}~\bibnamefont
				{Wells}}, \bibinfo {author} {\bibfnamefont {S.}~\bibnamefont {Kalliakos}},
			\bibinfo {author} {\bibfnamefont {B.}~\bibnamefont {Villa}}, \bibinfo
			{author} {\bibfnamefont {D.}~\bibnamefont {Ellis}}, \bibinfo {author}
			{\bibfnamefont {R.}~\bibnamefont {Stevenson}}, \bibinfo {author}
			{\bibfnamefont {A.}~\bibnamefont {Bennett}}, \bibinfo {author} {\bibfnamefont
				{I.}~\bibnamefont {Farrer}}, \bibinfo {author} {\bibfnamefont
				{D.}~\bibnamefont {Ritchie}},\ and\ \bibinfo {author} {\bibfnamefont
				{A.}~\bibnamefont {Shields}},\ }\bibfield  {title} {\bibinfo {title} {{Photon
					phase shift at the few-photon level and optical switching by a QD in a
					microcavity}},\ }\href@noop {} {\bibfield  {journal} {\bibinfo  {journal}
				{{Phys. Rev. A}}\ }\textbf {\bibinfo {volume} {11}},\ \bibinfo {pages}
			{061001} (\bibinfo {year} {2019})}\BibitemShut {NoStop}%
		\bibitem [{\citenamefont {Nowak}\ \emph {et~al.}(2014)\citenamefont {Nowak},
			\citenamefont {Portalupi}, \citenamefont {Giesz}, \citenamefont {Gazzano},
			\citenamefont {Dal~Savio}, \citenamefont {Braun}, \citenamefont {Karrai},
			\citenamefont {Arnold}, \citenamefont {Lanco}, \citenamefont {Sagnes},
			\citenamefont {Lemaître},\ and\ \citenamefont {Senellart}}]{Nowak2014}%
		\BibitemOpen
		\bibfield  {author} {\bibinfo {author} {\bibfnamefont {A.~K.}\ \bibnamefont
				{Nowak}}, \bibinfo {author} {\bibfnamefont {S.~L.}\ \bibnamefont
				{Portalupi}}, \bibinfo {author} {\bibfnamefont {V.}~\bibnamefont {Giesz}},
			\bibinfo {author} {\bibfnamefont {O.}~\bibnamefont {Gazzano}}, \bibinfo
			{author} {\bibfnamefont {C.}~\bibnamefont {Dal~Savio}}, \bibinfo {author}
			{\bibfnamefont {P.-F.}\ \bibnamefont {Braun}}, \bibinfo {author}
			{\bibfnamefont {K.}~\bibnamefont {Karrai}}, \bibinfo {author} {\bibfnamefont
				{C.}~\bibnamefont {Arnold}}, \bibinfo {author} {\bibfnamefont
				{L.}~\bibnamefont {Lanco}}, \bibinfo {author} {\bibfnamefont
				{I.}~\bibnamefont {Sagnes}}, \bibinfo {author} {\bibfnamefont
				{A.}~\bibnamefont {Lemaître}},\ and\ \bibinfo {author} {\bibfnamefont
				{P.}~\bibnamefont {Senellart}},\ }\bibfield  {title} {\bibinfo {title}
			{{Deterministic and electrically tunable bright single-photon source}},\
		}\href@noop {} {\bibfield  {journal} {\bibinfo  {journal} {{Nature Comm}}\
			}\textbf {\bibinfo {volume} {5}},\ \bibinfo {pages} {3240} (\bibinfo {year}
			{2014})}\BibitemShut {NoStop}%
		\bibitem [{\citenamefont {Hilaire}\ \emph {et~al.}(2018)\citenamefont
			{Hilaire}, \citenamefont {Antón}, \citenamefont {Kessler}, \citenamefont
			{Lemaître}, \citenamefont {Sagnes}, \citenamefont {Somaschi}, \citenamefont
			{Senellart},\ and\ \citenamefont {Lanco}}]{Hilaire2018}%
		\BibitemOpen
		\bibfield  {author} {\bibinfo {author} {\bibfnamefont {P.}~\bibnamefont
				{Hilaire}}, \bibinfo {author} {\bibfnamefont {C.}~\bibnamefont {Antón}},
			\bibinfo {author} {\bibfnamefont {C.}~\bibnamefont {Kessler}}, \bibinfo
			{author} {\bibfnamefont {A.}~\bibnamefont {Lemaître}}, \bibinfo {author}
			{\bibfnamefont {I.}~\bibnamefont {Sagnes}}, \bibinfo {author} {\bibfnamefont
				{N.}~\bibnamefont {Somaschi}}, \bibinfo {author} {\bibfnamefont
				{P.}~\bibnamefont {Senellart}},\ and\ \bibinfo {author} {\bibfnamefont
				{L.}~\bibnamefont {Lanco}},\ }\bibfield  {title} {\bibinfo {title} {{Accurate
					measurement of a $96\%$ input coupling into a cavity using polarization
					tomography}},\ }\href@noop {} {\bibfield  {journal} {\bibinfo  {journal}
				{{Appl. Phys. Lett.}}\ }\textbf {\bibinfo {volume} {112}},\ \bibinfo {pages}
			{201101} (\bibinfo {year} {2018})}\BibitemShut {NoStop}%
		\bibitem [{\citenamefont {Somaschi}\ \emph {et~al.}(2016)\citenamefont
			{Somaschi}, \citenamefont {Giesz}, \citenamefont {De~Santis}, \citenamefont
			{Loredo}, \citenamefont {Almeida}, \citenamefont {Hornecker}, \citenamefont
			{Portalupi}, \citenamefont {Grange}, \citenamefont {Antón}, \citenamefont
			{Demory}, \citenamefont {Gómez}, \citenamefont {Sagnes}, \citenamefont
			{Lanzillotti-Kimura}, \citenamefont {Lemaítre}, \citenamefont {Auffeves},
			\citenamefont {White}, \citenamefont {Lanco},\ and\ \citenamefont
			{Senellart}}]{Somaschi2016}%
		\BibitemOpen
		\bibfield  {author} {\bibinfo {author} {\bibfnamefont {N.}~\bibnamefont
				{Somaschi}}, \bibinfo {author} {\bibfnamefont {V.}~\bibnamefont {Giesz}},
			\bibinfo {author} {\bibfnamefont {L.}~\bibnamefont {De~Santis}}, \bibinfo
			{author} {\bibfnamefont {J.~C.}\ \bibnamefont {Loredo}}, \bibinfo {author}
			{\bibfnamefont {M.~P.}\ \bibnamefont {Almeida}}, \bibinfo {author}
			{\bibfnamefont {G.}~\bibnamefont {Hornecker}}, \bibinfo {author}
			{\bibfnamefont {S.~L.}\ \bibnamefont {Portalupi}}, \bibinfo {author}
			{\bibfnamefont {T.}~\bibnamefont {Grange}}, \bibinfo {author} {\bibfnamefont
				{C.}~\bibnamefont {Antón}}, \bibinfo {author} {\bibfnamefont
				{J.}~\bibnamefont {Demory}}, \bibinfo {author} {\bibfnamefont
				{C.}~\bibnamefont {Gómez}}, \bibinfo {author} {\bibfnamefont
				{I.}~\bibnamefont {Sagnes}}, \bibinfo {author} {\bibfnamefont {N.~D.}\
				\bibnamefont {Lanzillotti-Kimura}}, \bibinfo {author} {\bibfnamefont
				{A.}~\bibnamefont {Lemaítre}}, \bibinfo {author} {\bibfnamefont
				{A.}~\bibnamefont {Auffeves}}, \bibinfo {author} {\bibfnamefont {A.~G.}\
				\bibnamefont {White}}, \bibinfo {author} {\bibfnamefont {L.}~\bibnamefont
				{Lanco}},\ and\ \bibinfo {author} {\bibfnamefont {P.}~\bibnamefont
				{Senellart}},\ }\bibfield  {title} {\bibinfo {title} {{Near-optimal single
					photon sources in the solid state}},\ }\href@noop {} {\bibfield  {journal}
			{\bibinfo  {journal} {{Nature Photon.}}\ }\textbf {\bibinfo {volume} {10}},\
			\bibinfo {pages} {340} (\bibinfo {year} {2016})}\BibitemShut {NoStop}%
		\bibitem [{\citenamefont {Ding}\ \emph {et~al.}(2016)\citenamefont {Ding},
			\citenamefont {He}, \citenamefont {Duan}, \citenamefont {Gregersen},
			\citenamefont {Chen}, \citenamefont {Unsleber}, \citenamefont {Maier},
			\citenamefont {Schneider}, \citenamefont {Kamp}, \citenamefont {Höfling},
			\citenamefont {Lu},\ and\ \citenamefont {Pan}}]{Ding2016}%
		\BibitemOpen
		\bibfield  {author} {\bibinfo {author} {\bibfnamefont {X.}~\bibnamefont
				{Ding}}, \bibinfo {author} {\bibfnamefont {Y.}~\bibnamefont {He}}, \bibinfo
			{author} {\bibfnamefont {Z.-C.}\ \bibnamefont {Duan}}, \bibinfo {author}
			{\bibfnamefont {N.}~\bibnamefont {Gregersen}}, \bibinfo {author}
			{\bibfnamefont {M.-C.}\ \bibnamefont {Chen}}, \bibinfo {author}
			{\bibfnamefont {S.}~\bibnamefont {Unsleber}}, \bibinfo {author}
			{\bibfnamefont {S.}~\bibnamefont {Maier}}, \bibinfo {author} {\bibfnamefont
				{C.}~\bibnamefont {Schneider}}, \bibinfo {author} {\bibfnamefont
				{M.}~\bibnamefont {Kamp}}, \bibinfo {author} {\bibfnamefont {S.}~\bibnamefont
				{Höfling}}, \bibinfo {author} {\bibfnamefont {C.-Y.}\ \bibnamefont {Lu}},\
			and\ \bibinfo {author} {\bibfnamefont {J.-W.}\ \bibnamefont {Pan}},\
		}\bibfield  {title} {\bibinfo {title} {{On-Demand Single Photons with High
					Extraction Efficiency and Near-Unity Indistinguishability from a Resonantly
					Driven Quantum Dot in a Micropillar}},\ }\href@noop {} {\bibfield  {journal}
			{\bibinfo  {journal} {{Phys. Rev. Lett.}}\ }\textbf {\bibinfo {volume}
				{116}},\ \bibinfo {pages} {020401} (\bibinfo {year} {2016})}\BibitemShut
		{NoStop}%
		\bibitem [{\citenamefont {De~Santis}\ \emph {et~al.}(2017)\citenamefont
			{De~Santis}, \citenamefont {Antón}, \citenamefont {Reznychenko},
			\citenamefont {Somaschi}, \citenamefont {Coppola}, \citenamefont {Senellart},
			\citenamefont {Gómez}, \citenamefont {Lemaître}, \citenamefont {Sagnes},
			\citenamefont {White}, \citenamefont {Lanco}, \citenamefont {Auffèves},\
			and\ \citenamefont {Senellart}}]{DeSantis2017}%
		\BibitemOpen
		\bibfield  {author} {\bibinfo {author} {\bibfnamefont {L.}~\bibnamefont
				{De~Santis}}, \bibinfo {author} {\bibfnamefont {C.}~\bibnamefont {Antón}},
			\bibinfo {author} {\bibfnamefont {B.}~\bibnamefont {Reznychenko}}, \bibinfo
			{author} {\bibfnamefont {N.}~\bibnamefont {Somaschi}}, \bibinfo {author}
			{\bibfnamefont {G.}~\bibnamefont {Coppola}}, \bibinfo {author} {\bibfnamefont
				{J.}~\bibnamefont {Senellart}}, \bibinfo {author} {\bibfnamefont
				{C.}~\bibnamefont {Gómez}}, \bibinfo {author} {\bibfnamefont
				{A.}~\bibnamefont {Lemaître}}, \bibinfo {author} {\bibfnamefont
				{I.}~\bibnamefont {Sagnes}}, \bibinfo {author} {\bibfnamefont {A.~G.}\
				\bibnamefont {White}}, \bibinfo {author} {\bibfnamefont {L.}~\bibnamefont
				{Lanco}}, \bibinfo {author} {\bibfnamefont {A.}~\bibnamefont {Auffèves}},\
			and\ \bibinfo {author} {\bibfnamefont {P.}~\bibnamefont {Senellart}},\
		}\bibfield  {title} {\bibinfo {title} {{A solid-state single-photon
					filter}},\ }\href@noop {} {\bibfield  {journal} {\bibinfo  {journal} {{Nature
						Nano.}}\ }\textbf {\bibinfo {volume} {12}},\ \bibinfo {pages} {663} (\bibinfo
			{year} {2017})}\BibitemShut {NoStop}%
		\bibitem [{\citenamefont {Antón}\ \emph {et~al.}(2017)\citenamefont {Antón},
			\citenamefont {Hilaire}, \citenamefont {Kessler}, \citenamefont {Demory},
			\citenamefont {Gómez}, \citenamefont {Lemaître}, \citenamefont {Sagnes},
			\citenamefont {Lanzillotti-Kimura}, \citenamefont {Krebs}, \citenamefont
			{Somaschi}, \citenamefont {Senellart},\ and\ \citenamefont
			{Lanco}}]{Anton2017}%
		\BibitemOpen
		\bibfield  {author} {\bibinfo {author} {\bibfnamefont {C.}~\bibnamefont
				{Antón}}, \bibinfo {author} {\bibfnamefont {P.}~\bibnamefont {Hilaire}},
			\bibinfo {author} {\bibfnamefont {C.~A.}\ \bibnamefont {Kessler}}, \bibinfo
			{author} {\bibfnamefont {J.}~\bibnamefont {Demory}}, \bibinfo {author}
			{\bibfnamefont {C.}~\bibnamefont {Gómez}}, \bibinfo {author} {\bibfnamefont
				{A.}~\bibnamefont {Lemaître}}, \bibinfo {author} {\bibfnamefont
				{I.}~\bibnamefont {Sagnes}}, \bibinfo {author} {\bibfnamefont {N.~D.}\
				\bibnamefont {Lanzillotti-Kimura}}, \bibinfo {author} {\bibfnamefont
				{O.}~\bibnamefont {Krebs}}, \bibinfo {author} {\bibfnamefont
				{N.}~\bibnamefont {Somaschi}}, \bibinfo {author} {\bibfnamefont
				{P.}~\bibnamefont {Senellart}},\ and\ \bibinfo {author} {\bibfnamefont
				{L.}~\bibnamefont {Lanco}},\ }\bibfield  {title} {\bibinfo {title}
			{{Tomography of the optical polarization rotation induced by a single quantum
					dot in a cavity}},\ }\href@noop {} {\bibfield  {journal} {\bibinfo  {journal}
				{{Optica}}\ }\textbf {\bibinfo {volume} {4}},\ \bibinfo {pages} {1326}
			(\bibinfo {year} {2017})}\BibitemShut {NoStop}%
		\bibitem [{\citenamefont {Hilaire}\ \emph {et~al.}(2020)\citenamefont
			{Hilaire}, \citenamefont {Millet}, \citenamefont {Loredo}, \citenamefont
			{Antón}, \citenamefont {Harouri}, \citenamefont {Lemaître}, \citenamefont
			{Sagnes}, \citenamefont {Somaschi}, \citenamefont {Krebs}, \citenamefont
			{Senellart},\ and\ \citenamefont {Lanco}}]{Hilaire2020}%
		\BibitemOpen
		\bibfield  {author} {\bibinfo {author} {\bibfnamefont {P.}~\bibnamefont
				{Hilaire}}, \bibinfo {author} {\bibfnamefont {C.}~\bibnamefont {Millet}},
			\bibinfo {author} {\bibfnamefont {J.~C.}\ \bibnamefont {Loredo}}, \bibinfo
			{author} {\bibfnamefont {C.}~\bibnamefont {Antón}}, \bibinfo {author}
			{\bibfnamefont {A.}~\bibnamefont {Harouri}}, \bibinfo {author} {\bibfnamefont
				{A.}~\bibnamefont {Lemaître}}, \bibinfo {author} {\bibfnamefont
				{I.}~\bibnamefont {Sagnes}}, \bibinfo {author} {\bibfnamefont
				{N.}~\bibnamefont {Somaschi}}, \bibinfo {author} {\bibfnamefont
				{O.}~\bibnamefont {Krebs}}, \bibinfo {author} {\bibfnamefont
				{P.}~\bibnamefont {Senellart}},\ and\ \bibinfo {author} {\bibfnamefont
				{L.}~\bibnamefont {Lanco}},\ }\bibfield  {title} {\bibinfo {title}
			{{Deterministic assembly of a charged-quantum-dot-pillar cavity device}},\
		}\href@noop {} {\bibfield  {journal} {\bibinfo  {journal} {{Phys. Rev. B }}\
			}\textbf {\bibinfo {volume} {102}},\ \bibinfo {pages} {195402} (\bibinfo
			{year} {2020})}\BibitemShut {NoStop}%
		\bibitem [{\citenamefont {Warburton}(2013)}]{Warburton2013}%
		\BibitemOpen
		\bibfield  {author} {\bibinfo {author} {\bibfnamefont {R.~J.}\ \bibnamefont
				{Warburton}},\ }\bibfield  {title} {\bibinfo {title} {{Single spins in
					self-assembled quantum dots}},\ }\href@noop {} {\bibfield  {journal}
			{\bibinfo  {journal} {{Nature Materials}}\ }\textbf {\bibinfo {volume}
				{12}},\ \bibinfo {pages} {483} (\bibinfo {year} {2013})}\BibitemShut
		{NoStop}%
		\bibitem [{\citenamefont {Mannel}\ \emph {et~al.}(2021)\citenamefont {Mannel},
			\citenamefont {Kerski}, \citenamefont {Lochner}, \citenamefont {Zöllner},
			\citenamefont {Wieck}, \citenamefont {Ludwig}, \citenamefont {Lorke},\ and\
			\citenamefont {Geller}}]{Mannel2021}%
		\BibitemOpen
		\bibfield  {author} {\bibinfo {author} {\bibfnamefont {H.}~\bibnamefont
				{Mannel}}, \bibinfo {author} {\bibfnamefont {J.}~\bibnamefont {Kerski}},
			\bibinfo {author} {\bibfnamefont {P.}~\bibnamefont {Lochner}}, \bibinfo
			{author} {\bibfnamefont {M.}~\bibnamefont {Zöllner}}, \bibinfo {author}
			{\bibfnamefont {A.~D.}\ \bibnamefont {Wieck}}, \bibinfo {author}
			{\bibfnamefont {A.}~\bibnamefont {Ludwig}}, \bibinfo {author} {\bibfnamefont
				{A.}~\bibnamefont {Lorke}},\ and\ \bibinfo {author} {\bibfnamefont
				{M.}~\bibnamefont {Geller}},\ }\bibfield  {title} {\bibinfo {title} {{Auger
					and spin dynamics in a self-assembled quantum dot}},\ }\href@noop {}
		{\bibfield  {journal} {\bibinfo  {journal} {{arXiv:2110.12213}}\ } (\bibinfo
			{year} {2021})}\BibitemShut {NoStop}%
		\bibitem [{\citenamefont {Urbaszek}\ \emph {et~al.}(2013)\citenamefont
			{Urbaszek}, \citenamefont {Marie}, \citenamefont {Amand}, \citenamefont
			{Krebs}, \citenamefont {Voisin}, \citenamefont {Maletinsky}, \citenamefont
			{Högele},\ and\ \citenamefont {Imamoglu}}]{Urbaszek2013}%
		\BibitemOpen
		\bibfield  {author} {\bibinfo {author} {\bibfnamefont {B.}~\bibnamefont
				{Urbaszek}}, \bibinfo {author} {\bibfnamefont {X.}~\bibnamefont {Marie}},
			\bibinfo {author} {\bibfnamefont {T.}~\bibnamefont {Amand}}, \bibinfo
			{author} {\bibfnamefont {O.}~\bibnamefont {Krebs}}, \bibinfo {author}
			{\bibfnamefont {P.}~\bibnamefont {Voisin}}, \bibinfo {author} {\bibfnamefont
				{P.}~\bibnamefont {Maletinsky}}, \bibinfo {author} {\bibfnamefont
				{A.}~\bibnamefont {Högele}},\ and\ \bibinfo {author} {\bibfnamefont
				{A.}~\bibnamefont {Imamoglu}},\ }\bibfield  {title} {\bibinfo {title}
			{{Nuclear spin physics in quantum dots: An optical investigation}},\
		}\href@noop {} {\bibfield  {journal} {\bibinfo  {journal} {{Rev. Mod. Phys}}\
			}\textbf {\bibinfo {volume} {7}},\ \bibinfo {pages} {79} (\bibinfo {year}
			{2013})}\BibitemShut {NoStop}%
		\bibitem [{\citenamefont {Tan}(1999)}]{Tan1999}%
		\BibitemOpen
		\bibfield  {author} {\bibinfo {author} {\bibfnamefont {S.~M.}\ \bibnamefont
				{Tan}},\ }\bibfield  {title} {\bibinfo {title} {{A computational toolbox for
					quantum and atomic optics}},\ }\href@noop {} {\bibfield  {journal} {\bibinfo
				{journal} {{J. Opt. B: Quantum Semiclass. Opt}}\ }\textbf {\bibinfo {volume}
				{1}},\ \bibinfo {pages} {424} (\bibinfo {year} {1999})}\BibitemShut {NoStop}%
		\bibitem [{\citenamefont {Zhukov}\ \emph {et~al.}(2018)\citenamefont {Zhukov},
			\citenamefont {Kirstein}, \citenamefont {Smirnov}, \citenamefont {Yakovlev},
			\citenamefont {Glazov}, \citenamefont {Reuter}, \citenamefont {Wieck},
			\citenamefont {Bayer},\ and\ \citenamefont {Greilich}}]{Zhukov2018}%
		\BibitemOpen
		\bibfield  {author} {\bibinfo {author} {\bibfnamefont {E.~A.}\ \bibnamefont
				{Zhukov}}, \bibinfo {author} {\bibfnamefont {E.}~\bibnamefont {Kirstein}},
			\bibinfo {author} {\bibfnamefont {D.~S.}\ \bibnamefont {Smirnov}}, \bibinfo
			{author} {\bibfnamefont {D.~R.}\ \bibnamefont {Yakovlev}}, \bibinfo {author}
			{\bibfnamefont {M.~M.}\ \bibnamefont {Glazov}}, \bibinfo {author}
			{\bibfnamefont {D.}~\bibnamefont {Reuter}}, \bibinfo {author} {\bibfnamefont
				{A.~D.}\ \bibnamefont {Wieck}}, \bibinfo {author} {\bibfnamefont
				{M.}~\bibnamefont {Bayer}},\ and\ \bibinfo {author} {\bibfnamefont
				{A.}~\bibnamefont {Greilich}},\ }\bibfield  {title} {\bibinfo {title} {{Spin
					inertia of resident and photoexcited carriers in singly charged quantum
					dots}},\ }\href@noop {} {\bibfield  {journal} {\bibinfo  {journal} {{Phys.
						Rev. B}}\ }\textbf {\bibinfo {volume} {98}},\ \bibinfo {pages} {121304}
			(\bibinfo {year} {2018})}\BibitemShut {NoStop}%
		\bibitem [{\citenamefont {Kuhlmann}\ \emph {et~al.}(2013)\citenamefont
			{Kuhlmann}, \citenamefont {Houel}, \citenamefont {Ludwig}, \citenamefont
			{Greuter}, \citenamefont {Reuter}, \citenamefont {Wieck}, \citenamefont
			{Poggio},\ and\ \citenamefont {Warburton}}]{Kuhlmann2013}%
		\BibitemOpen
		\bibfield  {author} {\bibinfo {author} {\bibfnamefont {A.~V.}\ \bibnamefont
				{Kuhlmann}}, \bibinfo {author} {\bibfnamefont {J.}~\bibnamefont {Houel}},
			\bibinfo {author} {\bibfnamefont {A.}~\bibnamefont {Ludwig}}, \bibinfo
			{author} {\bibfnamefont {L.}~\bibnamefont {Greuter}}, \bibinfo {author}
			{\bibfnamefont {D.}~\bibnamefont {Reuter}}, \bibinfo {author} {\bibfnamefont
				{A.~D.}\ \bibnamefont {Wieck}}, \bibinfo {author} {\bibfnamefont
				{M.}~\bibnamefont {Poggio}},\ and\ \bibinfo {author} {\bibfnamefont {R.~J.}\
				\bibnamefont {Warburton}},\ }\bibfield  {title} {\bibinfo {title} {{Charge
					noise and spin noise in a semiconductor quantum device}},\ }\href@noop {}
		{\bibfield  {journal} {\bibinfo  {journal} {{Nature Physics}}\ }\textbf
			{\bibinfo {volume} {85}},\ \bibinfo {pages} {570} (\bibinfo {year}
			{2013})}\BibitemShut {NoStop}%
		\bibitem [{\citenamefont {{Altepeter, Joseph B. and James, Daniel F.V. and
					Kwiat, Paul G.}}(2004)}]{Kwiat2004}%
		\BibitemOpen
		\bibfield  {author} {\bibinfo {author} {\bibnamefont {{Altepeter, Joseph B.
						and James, Daniel F.V. and Kwiat, Paul G.}}},\ }\href@noop {} {\emph
			{\bibinfo {title} {{Quantum State Estimation}}}}\ (\bibinfo  {publisher}
		{{Springer}},\ \bibinfo {year} {2004})\ Chap.~\bibinfo {chapter}
		{4}\BibitemShut {NoStop}%
		\bibitem [{\citenamefont {Prechtel}\ \emph {et~al.}(2016)\citenamefont
			{Prechtel}, \citenamefont {Kuhlmann}, \citenamefont {Houel}, \citenamefont
			{Ludwig}, \citenamefont {Valentin}, \citenamefont {Wieck},\ and\
			\citenamefont {Warburton}}]{Prechtel2016}%
		\BibitemOpen
		\bibfield  {author} {\bibinfo {author} {\bibfnamefont {J.~H.}\ \bibnamefont
				{Prechtel}}, \bibinfo {author} {\bibfnamefont {A.~V.}\ \bibnamefont
				{Kuhlmann}}, \bibinfo {author} {\bibfnamefont {J.}~\bibnamefont {Houel}},
			\bibinfo {author} {\bibfnamefont {A.}~\bibnamefont {Ludwig}}, \bibinfo
			{author} {\bibfnamefont {S.~R.}\ \bibnamefont {Valentin}}, \bibinfo {author}
			{\bibfnamefont {A.~D.}\ \bibnamefont {Wieck}},\ and\ \bibinfo {author}
			{\bibfnamefont {R.~J.}\ \bibnamefont {Warburton}},\ }\bibfield  {title}
		{\bibinfo {title} {{Decoupling a hole spin qubit from the nuclear spins}},\
		}\href@noop {} {\bibfield  {journal} {\bibinfo  {journal} {{Nature
						Materials}}\ }\textbf {\bibinfo {volume} {15}},\ \bibinfo {pages} {981}
			(\bibinfo {year} {2016})}\BibitemShut {NoStop}%
		\bibitem [{\citenamefont {Gangloff}\ \emph {et~al.}(2019)\citenamefont
			{Gangloff}, \citenamefont {Éthier Majcher}, \citenamefont {Lang},
			\citenamefont {Denning}, \citenamefont {Bodey}, \citenamefont {Jackson},
			\citenamefont {Clarke}, \citenamefont {Hugues}, \citenamefont {Le~Gall},\
			and\ \citenamefont {Atatüre}}]{Gangloff2019}%
		\BibitemOpen
		\bibfield  {author} {\bibinfo {author} {\bibfnamefont {D.~A.}\ \bibnamefont
				{Gangloff}}, \bibinfo {author} {\bibfnamefont {G.}~\bibnamefont {Éthier
					Majcher}}, \bibinfo {author} {\bibfnamefont {C.}~\bibnamefont {Lang}},
			\bibinfo {author} {\bibfnamefont {E.~V.}\ \bibnamefont {Denning}}, \bibinfo
			{author} {\bibfnamefont {J.~H.}\ \bibnamefont {Bodey}}, \bibinfo {author}
			{\bibfnamefont {D.~M.}\ \bibnamefont {Jackson}}, \bibinfo {author}
			{\bibfnamefont {E.}~\bibnamefont {Clarke}}, \bibinfo {author} {\bibfnamefont
				{M.}~\bibnamefont {Hugues}}, \bibinfo {author} {\bibfnamefont
				{C.}~\bibnamefont {Le~Gall}},\ and\ \bibinfo {author} {\bibfnamefont
				{M.}~\bibnamefont {Atatüre}},\ }\bibfield  {title} {\bibinfo {title}
			{{Quantum interface of an electron and a nuclear ensemble}},\ }\href@noop {}
		{\bibfield  {journal} {\bibinfo  {journal} {{Science}}\ }\textbf {\bibinfo
				{volume} {364}},\ \bibinfo {pages} {62} (\bibinfo {year} {2019})}\BibitemShut
		{NoStop}%
		\bibitem [{\citenamefont {Singh}\ \emph {et~al.}(2022)\citenamefont {Singh},
			\citenamefont {Farfurnik}, \citenamefont {Luo}, \citenamefont {Bracker},
			\citenamefont {Carter},\ and\ \citenamefont {Waks}}]{Singh2022}%
		\BibitemOpen
		\bibfield  {author} {\bibinfo {author} {\bibfnamefont {H.}~\bibnamefont
				{Singh}}, \bibinfo {author} {\bibfnamefont {D.}~\bibnamefont {Farfurnik}},
			\bibinfo {author} {\bibfnamefont {Z.}~\bibnamefont {Luo}}, \bibinfo {author}
			{\bibfnamefont {A.~S.}\ \bibnamefont {Bracker}}, \bibinfo {author}
			{\bibfnamefont {S.~G.}\ \bibnamefont {Carter}},\ and\ \bibinfo {author}
			{\bibfnamefont {E.}~\bibnamefont {Waks}},\ }\bibfield  {title} {\bibinfo
			{title} {{Optical Transparency Induced by a Largely Purcell Enhanced Quantum
					Dot in a Polarization-Degenerate Cavity}},\ }\href@noop {} {\bibfield
			{journal} {\bibinfo  {journal} {{Nano Lett.}}\ }\textbf {\bibinfo {volume}
				{22}},\ \bibinfo {pages} {7959} (\bibinfo {year} {2022})}\BibitemShut
		{NoStop}%
		\bibitem [{\citenamefont {Luxmoore}\ \emph {et~al.}(2013)\citenamefont
			{Luxmoore}, \citenamefont {Wasley}, \citenamefont {Ramsay}, \citenamefont
			{Thijssen}, \citenamefont {Oulton}, \citenamefont {Hugues}, \citenamefont
			{Kasture}, \citenamefont {Achanta}, \citenamefont {Fox},\ and\ \citenamefont
			{Skolnick}}]{Luxmoore2013}%
		\BibitemOpen
		\bibfield  {author} {\bibinfo {author} {\bibfnamefont {I.~J.}\ \bibnamefont
				{Luxmoore}}, \bibinfo {author} {\bibfnamefont {N.~A.}\ \bibnamefont
				{Wasley}}, \bibinfo {author} {\bibfnamefont {A.~J.}\ \bibnamefont {Ramsay}},
			\bibinfo {author} {\bibfnamefont {A.~C.~T.}\ \bibnamefont {Thijssen}},
			\bibinfo {author} {\bibfnamefont {R.}~\bibnamefont {Oulton}}, \bibinfo
			{author} {\bibfnamefont {M.}~\bibnamefont {Hugues}}, \bibinfo {author}
			{\bibfnamefont {S.}~\bibnamefont {Kasture}}, \bibinfo {author} {\bibfnamefont
				{V.~G.}\ \bibnamefont {Achanta}}, \bibinfo {author} {\bibfnamefont {A.~M.}\
				\bibnamefont {Fox}},\ and\ \bibinfo {author} {\bibfnamefont {M.~S.}\
				\bibnamefont {Skolnick}},\ }\bibfield  {title} {\bibinfo {title}
			{{Interfacing Spins in an InGaAs Quantum Dot to a Semiconductor Waveguide
					Circuit Using Emitted Photons}},\ }\href@noop {} {\bibfield  {journal}
			{\bibinfo  {journal} {{Phys. Rev. Lett.}}\ }\textbf {\bibinfo {volume}
				{110}},\ \bibinfo {pages} {037402} (\bibinfo {year} {2013})}\BibitemShut
		{NoStop}%
		\bibitem [{\citenamefont {Javadi}\ \emph {et~al.}(2018)\citenamefont {Javadi},
			\citenamefont {Ding}, \citenamefont {Appel}, \citenamefont {Mahmoodian},
			\citenamefont {Löbl}, \citenamefont {Söllner}, \citenamefont {Schott},
			\citenamefont {Papon}, \citenamefont {Pregnolato}, \citenamefont {Stobbe},
			\citenamefont {Midolo}, \citenamefont {Schröder}, \citenamefont {Wieck},
			\citenamefont {Ludwig}, \citenamefont {Warburton},\ and\ \citenamefont
			{Lodahl}}]{Javadi2018}%
		\BibitemOpen
		\bibfield  {author} {\bibinfo {author} {\bibfnamefont {A.}~\bibnamefont
				{Javadi}}, \bibinfo {author} {\bibfnamefont {D.}~\bibnamefont {Ding}},
			\bibinfo {author} {\bibfnamefont {M.~H.}\ \bibnamefont {Appel}}, \bibinfo
			{author} {\bibfnamefont {S.}~\bibnamefont {Mahmoodian}}, \bibinfo {author}
			{\bibfnamefont {M.~C.}\ \bibnamefont {Löbl}}, \bibinfo {author}
			{\bibfnamefont {I.}~\bibnamefont {Söllner}}, \bibinfo {author}
			{\bibfnamefont {R.}~\bibnamefont {Schott}}, \bibinfo {author} {\bibfnamefont
				{C.}~\bibnamefont {Papon}}, \bibinfo {author} {\bibfnamefont
				{T.}~\bibnamefont {Pregnolato}}, \bibinfo {author} {\bibfnamefont
				{S.}~\bibnamefont {Stobbe}}, \bibinfo {author} {\bibfnamefont
				{L.}~\bibnamefont {Midolo}}, \bibinfo {author} {\bibfnamefont
				{T.}~\bibnamefont {Schröder}}, \bibinfo {author} {\bibfnamefont {A.~D.}\
				\bibnamefont {Wieck}}, \bibinfo {author} {\bibfnamefont {A.}~\bibnamefont
				{Ludwig}}, \bibinfo {author} {\bibfnamefont {R.~J.}\ \bibnamefont
				{Warburton}},\ and\ \bibinfo {author} {\bibfnamefont {P.}~\bibnamefont
				{Lodahl}},\ }\bibfield  {title} {\bibinfo {title} {{Spin–photon interface
					and spin-controlled photon switching in a nanobeam waveguide}},\ }\href@noop
		{} {\bibfield  {journal} {\bibinfo  {journal} {{Nature Nano}}\ }\textbf
			{\bibinfo {volume} {13}},\ \bibinfo {pages} {398–403} (\bibinfo {year}
			{2018})}\BibitemShut {NoStop}%
	\end{thebibliography}
	
	%

	%
	%
	\section*{Acknowledgements}
	
	This work was partially supported by the Paris Ile-de-France Région in the framework of DIM SIRTEQ,
	the European Union’s Horizon 2020 Research and Innovation Programme QUDOT-TECH under the Marie Sklodowska-Curie Grant Agreement No. 861097,
	the European Union’s Horizon 2020 FET OPEN project QLUSTER (Grant ID 862035),
	the French RENATECH network and 
	a public grant overseen by the French National Research Agency (ANR) as part of the ”Investissements d’Avenir” programme (Labex NanoSaclay, reference: ANR-10-LABX-0035).\\

	\section*{Author contributions}
	
	E.M. and M.G.M. performed the experiments, with the help of C.M. who participated in the experimental developments. E.M. analysed the data with the help of M.G-M..
	A.L., I.S., N.S. and L.L.G. fabricated the device based on a design by P.S.. 
	C.M., E.M. and M.G-M developed the numerical simulations. 
	L.L. conducted the project. 
	All authors participated in scientific discussions and manuscript preparation.

\end{document}